# A Machine Learning Data Fusion Model for Soil Moisture Retrieval

Vishal Batchu[a], Grey Nearing[a], Varun Gulshan[a].

[a] *Google Research*

*Corresponding author*: Vishal Batchu, vishalbatchu@google.com

ABSTRACT

We develop a deep learning based convolutional-regression model that estimates the volumetric soil moisture content in the top ~5 cm of soil. Input predictors include Sentinel-1 (active radar), Sentinel-2 (optical imagery), and SMAP (passive radar) as well as geophysical variables from SoilGrids and modelled soil moisture fields from GLDAS. The model was trained and evaluated on data from ~1300 in-situ sensors globally over the period 2015 - 2021 and obtained an average per-sensor correlation of 0.727 and ubRMSE of 0.054, and can be used to produce a soil moisture map at a nominal 320m resolution. These results are benchmarked against 13 other soil moisture works at different locations, and an ablation study was used to identify important predictors.

SIGNIFICANCE STATEMENT

Soil moisture is a key variable in various agriculture and water management systems. Accurate and high resolution estimates of soil moisture have multiple downstream benefits such as reduced water wastage by better understanding and managing the consumption of water, utilising smarter irrigation methods and effective canal water management. We develop a deep learning based model that estimates the volumetric soil moisture content in the top ~5 cm of soil at a nominal 320m resolution. Our results demonstrate that machine learning is a useful tool for fusing different modalities with ease, while producing high resolution models that are not location specific. Future work could explore the possibility of using temporal input sources to further improve model performance.

## 1. Introduction

Soil moisture is one of the primary hydrological state (memory) variables in terrestrial systems (Dobriyal et al. 2012; Rossato et al. 2017a), and is one of the primary controls for agriculture and water management (Dobriyal et al. 2012; Rossato et al. 2017b). Soil moisture affects evapotranspiration and vegetation water availability, which are at the core of the climate-carbon cycle (Falloon et al. 2011) and play an important role in hydrological risks such as floods, drought, erosion, and landslides (Kim et al. 2019; Legates et al. 2011; Tramblay et al. 2012). Accurate measurement of soil moisture has numerous downstream benefits (Moran et al. 2015) including reduced water wastage by better understanding and managing the consumption of water (Brocca et al. 2018; Foster, Mieno, and Brozović 2020), utilising smarter irrigation methods (Kumar et al. 2014) and effective canal water management (Zafar, Prathapar, and Bastiaanssen 2021).

The most accurate way to measure soil moisture is via ground-based methods such as direct gravimetric measurements (Klute 1986) or indirect methods such as dielectric reflectometry, capacitance charge, etc. (Bittelli 2011), which in-situ sensors utilize (Walker, Willgoose, and Kalma 2004). However, in-situ sensors are difficult to scale spatially, and are expensive to install and maintain. Remote sensing-based methods scale globally and provide meaningfully accurate estimates of top-level (typically 0-5 cm) soil moisture (Ulaby, Moore, and Fung 1982), while lowering deployment and maintenance costs relative to ground based methods. While remote sensing soil moisture estimates generally have lower accuracy than in-situ measurements, they scale well spatially.

Existing remote sensing-based methods generally use microwave band radiometric reflectance to quantify soil moisture. These sensors can either be located on aerial or satellite platforms, and can be broadly classified into two types - passive and active.

Passive remote sensing mainly uses L/C-band brightness temperatures. Early retrieval methods were site specific, semi-empirical models such as Oh (Oh, Sarabandi, and Ulaby 1992), Dubois/Topp (Dubois, van Zyl, and Engman 1995) and the IEM models (Baghdadi et al. 2004; K. S. Chen et al. 2003). These methods provide reasonably accurate estimates of soil moisture, however, they are sensitive to site-specific parameters such as soil roughness (Mattia et al. 1997), and are only capable of estimating the soil moisture on bare-ground soils (Verhoest et al. 2008) or soils with low vegetation content. These models have been extensively tested and evaluated (MirMazloumi and Sahebi 2016; Panciera et al. 2014; Ma, Han, and Liu 2021; Choker et al. 2017), showing that they are generally not reliable when scaled globally. More generalizable retrieval methods were developed on top of remote sensing systems such as SMAP (Entekhabi et al. 2010), SMOS (Y. H. Kerr et al. 2001) and ASMR-E (Njoku et al. 2003). Baseline algorithms developed for each of these systems (Y. Kerr et al. 2012; Chan, Njoku, and Colliander 2012; Njoku et al. 2003) were physics-based models built on top of the microwave brightness temperatures validated with in-situ sensor grids (Al-Yaari et al. 2017; Cai et al. 2017; Q. Chen et al. 2018; Colliander et al. 2017). In addition to the brightness temperatures, these models also use various globally available static land surface parameters such as soil type, land cover, etc. (Chan, Njoku, and Colliander 2012). The main limitation of passive microwave products is their coarse spatial resolution, often at a resolution of 9 km x 9 km or more (Entekhabi et al. 2010) which is too coarse for many agriculture related tasks, e.g., field-scale monitoring.

Active remote sensing methods mainly utilise L/C-band Synthetic Aperture Radar (SAR), and allow for more high-resolution estimates (Haider et al. 2004; Shi et al. 1997). However, the retrieval accuracy of these methods is not as high as passive remote sensing methods (Njoku et al. 2002). Active and passive methods can be combined (Das et al. 2019; Wu et al. 2017; Das et al. 2018) to improve resolution and maintain accuracy. Passive and active methods are complementary – the former is more accurate but has a coarse spatial resolution while the latter is less accurate but has a higher spatial resolution – was an impetus for the SMAP satellite.

Recent work has looked at the combination of passive and active radar with other remote sensing sources, such as optical imagery (Ojha et al. 2021; Gao et al. 2017). A SMAP +

Sentinel-1 product was also developed from the perspective of disaggregating SMAP radiometer brightness temperatures based on Sentinel-1 readings (Das et al. 2019), however, this method has not yet achieved accuracies similar to SMAP at a higher resolution.

In recent years, the use of machine learning (ML) has shown promise in various fields of geoscience (Lary et al. 2016). ML has been used specifically to help disaggregate microwave-based soil moisture estimates into higher resolution products (Kolassa et al. 2018; Mao et al. 2019; Liu et al. 2020; Abbaszadeh, Moradkhani, and Zhan 2019). Machine learning provides a unique set of abilities that work well for soil-moisture estimation — enabling the use of large datasets such as the ISMN database (W. Dorigo et al. 2011) without the need for site-specific tuning, modelling non-linear relationships between multiple predictors (remote sensed inputs) and the target (SM), and fusing multiple sources of inputs which could potentially aid in handling vegetation and canopy covers (Lee et al. 2018). While some of these ML efforts have shown improvements in the estimation of soil moisture compared to previous methods, these techniques mostly depend on a small number of inputs (Liu et al. 2020) and face issues with speckle in active remote sensing data (Oliver and Quegan 1998) and are unable to perform scene understanding which is required in soil moisture estimation (Davenport, Sandells, and Gurney 2008). To address the aforementioned issues, we develop, train, and test a deep learning model that provides high resolution (nominal resolution of 320m - discussed in detail in Section 3e) and accurate (average Pearson correlation of 0.727) estimates of soil moisture globally

## 2. Data

We generate large datasets for training, evaluation, and testing.. Each data point consists of (i) a set of model inputs from various sources (described presently), (ii) soil moisture labels that the model is trained to estimate, and (iii) additional metadata such as timestamp, geographical coordinates etc. These data, their sources, and pre-processing are described in the following subsections.

*a. Labels*

1) INTERNATIONAL SOIL MOISTURE NETWORK

The largest repository (to our knowledge) of in-situ sensor data for soil moisture is from the ISMN 'network of networks' (W. Dorigo et al. 2011). This dataset has been used extensively for calibrating, training, and evaluating models, and we use it to maintain consistency with previous studies. At each sensor location, ISMN provides measurements in volumetric units at hourly intervals at various depths. The data are quality checked and flagged for anomalies/inconsistencies (W. A. Dorigo et al. 2013) in accordance with NASA's good validation practices (C. Montzka et al. 2020). We use ISMN soil moisture tagged with a depth of 0-5 cm and 5-5 cm. Both of these correspond to the top layer of soil moisture, however, the notation differs across different provider networks.

| Network | Countries | #Sensors |
| --- | --- | --- |
| AMMA-CATCH (Galle et al. 2018) | Benin, Niger, Mali | 4 |
| BIEBRZA_S-1 (Dąbrowska-Zielińska et al. 2018) | Poland | 25 |
| FR_Aqui (Alyaari et al. 2018) | France | 3 |
| IMA_CAN1 (Biddoccu et al. 2016) | Italy | 10 |
| MAQU (Su et al. 2011) | China | 15 |
| NAQU (Su et al. 2011) | China | 7 |
| REMEDHUS (González-Zamora et al. 2019) | Spain | 20 |
| RSMN (Sandric et al. 2016) | Romania | 19 |
| SCAN (Schaefer, Cosh, and Jackson 2007) | USA | 179 |
| SMOSMANIA (Calvet et al. 2016) | France | 10 |
| SNOTEL (Leavesley et al. 2010) | USA | 372 |
| SOILSCAPE | USA | 60 |
| USCRN (Bell et al. 2013) | USA | 114 |
| Total | | 838 |

Table 1. List of sensor networks from ISMN that we use as a part of our study.

The area of study is limited by the presence of in-situ sensors. Although the ISMN repository has a global presence, sensors are primarily located in the US, Europe, Australia and China (W. Dorigo et al. 2021). This limits the applicability of models trained on this data to similar geographies. The sensor networks that we use from ISMN are listed in Table 1.

2) SMAP Core Validation Sites

In addition to the ISMN network, we also train and validate our model using data from the SMAP core validation sites (Colliander et al. 2017) listed in Table 2.

| Network | Countries | # Sites/Sensors |
|---|---|---|
| FORTCOBB (Starks et al. 2014) | USA | 17 |
| KYEAMBA (Smith et al. 2012) | Australia | 8 |
| LITTLEWASHITA (Starks et al. 2014) | USA | 21 |
| TWENTE (Velde and Benninga 2021) | Netherlands | 22 |
| TXSON (Caldwell et al. 2019) | USA | 40 |
| YANCO (Smith et al. 2012) | Australia | 12 |
| Total | | 120 |

Table 2. List of SMAP core validation sites that we use as a part of our study. We only use a subset of all the core validation sites ("SMAP/In Situ Core Validation Site Land Surface Parameters Match-Up Data" n.d.) due to data access limitations.

*b. Input Data*

Input data for our models were downloaded from Google Earth Engine (Gorelick et al. 2017), which provides a number of different satellite products and geophysical variables that can be combined and exported in a format suitable for machine learning. Earth Engine also facilitates the processing of imagery such as scaling it to a given spatial resolution, performing temporal joins, projecting the imagery to a specific projection etc.

As mentioned in the introduction there are a number of types of remote sensing sources that can be useful for soil moisture estimation (Ahmed, Zhang, and Nichols 2011). We select sources that have a significant correlation with soil moisture and/or have potential to help with the disaggregation of low resolution soil moisture estimates.

1) High resolution sources

*(i) Sentinel-1 (S1)*

The Copernicus Sentinel-1 (Torres et al. 2012) mission (2014-present) by the European Space Agency (ESA) provides global Synthetic Aperture Radar (SAR) readings (Rosen et al. 2000) at regular intervals. It has a revisit time of 6 days at the equator (Torres et al. 2012).

We use the Sentinel-1 GRD product from Earth Engine which consists of VV, VH and angle imagery corresponding to dual-band cross-polarized data at a 10 meter resolution. The scenes undergo thermal noise removal, radiometric calibration and terrain correction with the Sentinel-1 toolbox (Veci et al. 2014) to despeckle and denoise.

*(ii) Sentinel-2 (S2)*

The Copernicus Sentinel-2 (Drusch et al. 2012) mission (2015-present) by the ESA provides high-resolution multispectral imagery (Table 3 lists the Sentinel-2 bands used in this study). It has a revisit time of 5 days at the equator.

Unlike SAR, optical, NIR, and SWIR imagery are dependent on cloud cover and the time of the day of acquisition. A large fraction of the Sentinel-2 scenes have significant cloud cover. We filter the data to retain only scenes containing less than 30% cloud cover (we use the QA60 cloud_mask band from Sentinel-2 to do this).

We use the L1C top of atmosphere product, which consists of multiple bands with varying resolutions ranging from 10-60m. All the bands were upscaled or maintained at a 10m resolution to use as inputs for our models.

| Band | Resolution | Wavelength | Description |
| --- | --- | --- | --- |
| B2 | 10 meters | 496.6nm (S2A) / 492.1nm (S2B) | Blue |
| B3 | 10 meters | 560nm (S2A) / 559nm (S2B) | Green |
| B4 | 10 meters | 664.5nm (S2A) / 665nm (S2B) | Red |
| B8 | 10 meters | 835.1nm (S2A) / 833nm (S2B) | NIR |
| B11 | 20 meters | 1613.7nm (S2A) / 1610.4nm (S2B) | SWIR 1 |
| B12 | 20 meters | 2202.4nm (S2A) / 2185.7nm (S2B) | SWIR 2 |

Table 3. Sentinel-2 bands we use and their wavelengths/properties.

*(iii) NASA Digital Elevation Model (DEM)*

Digital elevation models capture the topography of bare ground which helps estimate the amount of moisture the surface can hold. The NASA Digital Elevation Model (DEM) (Jpl 2020) provides 30 meter resolution estimates and is a reprocessing of the widely used SRTM product (Farr et al. 2000) which improves the model globally. Data from ASTER GDEM, ICESat GLAS and PRISM are incorporated into the SRTM product to produce the refined NASA DEM. This is a single time acquisition product (acquired in the year 2000) but the variation over time is small, hence the product is still relevant for the task at hand.

2) Low resolution sources

*(i) Soil Grids*

SoilGrids (Hengl et al. 2017) provides us with various environmental and soil profile layers with global coverage. We specifically use soil texture – i.e., sand, silt, and clay fractions - and bulk density. All of these mapped products are present at a resolution of 250m.

We do not use pedotransfer functions explicitly since our models can learn the required mapping. Soil information is essential because it is related to maximum water holding capacity (porosity), infiltration and to some extent, evaporation rates, which are critical controls on water retention and storage (Pan et al. 2012; Beale et al. 2019). Soil information obtained from soil grids has been used as a proxy for the disaggregation of soil moisture in prior work (Leenaars et al. 2018; Carsten Montzka et al. 2018).

3) Coarse soil moisture products

*(i) SMAP Soil Moisture*

The NASA USDA enhanced Soil Moisture Active Passive (SMAP) soil moisture product (Bolten et al. 2010) provides surface level (0-5 cm) soil moisture estimates at a 10km resolution. The product is produced by applying a Palmer model ("Meteorological Drought" n.d.) followed by 1-D Ensemble Kalman Filter (EnKF) (Evensen 2003) to assimilate the Level 3 SMAP product. It has a temporal revisit period of ~3 days.

The soil moisture provided by this product (ssm) is in units mm which is equivalent to kg / m^2. We convert this to volumetric soil moisture content by using 1000 kg / m^3 as the density of water and a measurement depth of 5cm = 0.05m[1].

*(ii) GLDAS Soil Moisture*

Global Land Data Assimilation System (GLDAS) (Rodell et al. 2004) uses land surface modelling and data assimilation techniques to model various land surface states and fluxes. They provide soil moisture products at various depths, of which 0-10 cm is closest to the depth that we want to estimate at (0-5 cm) at a 25 km resolution. Although their soil moisture product does not map exactly to the top level surface soil moisture, it still correlates well and is a useful input. Being a modelling product, GLDAS provides information indirectly from recent rainfall and meteorological inputs. Estimates are produced at a ~3 hr interval.

---

[1] *Additional information on the volumetric soil moisture conversion at* https://ldas.gsfc.nasa.gov/faq/

Similar to the SMAP data, we convert GLDAS SM provided in mm (kg/m^2) to volumetric soil moisture using a measurement depth of 10 cm.

4) INPUT NORMALIZATION

All inputs from Earth Engine data sources are normalized into a consistent range to prepare inputs for our machine learning model. A linear scaling (zero-mean, unit-variance) is used for most of the data sources.

| Source | Band | Min | Max |
|---|---|---|---|
| Sentinel-1 | VV | -25 | 5 |
| | VH | -25 | 5 |
| | Angle | 0 | 90 |
| MODIS LAI | All bands | 0 | 100 |
| Soil Grids | All bands | 0 | 800 |
| NASA DEM | elevation | 0 | 3000 |
| GLDAS | SoilMoi0_10cm_inst | 0 | 100 |
| SMAP | ssm | 50 | 255 |

Table 4. Source statistics for each of the Earth Engine sources we use. Each of these sources is normalized to zero-mean, unit-variance.

Coming to Sentinel-2, cloudy pixels in Sentinel-2 scenes have large reflectance values compared to non cloudy pixels. Due to this, a linear scaling across the entire data results in a small dynamic range for non-cloudy reflectance values. To account for this, we use a logarithm-based nonlinear scaling method (Brown et al. 2022) that results in better dynamic ranges for non-cloudy pixels. Additional details are specified in the Appendix.

*c. Creating the dataset*

Data from all the sources specified above are joined with a distributed Earth Engine export pipeline to create datasets for training and testing.

This pipeline consists of the following steps:
1. Filter ISMN dataset based on the quality flags. We retain data that has the following ISMN flags - "G", "C02", "C03", "C02,C03".
2. For each remaining ISMN data point consisting of latitude, longitude, timestamp, and soil moisture reading:
    a. For each earth engine source
        i. [Sentinel-2 source only] Filter out images with >30% cloud percentage.
        ii. Perform a spatial join to find matching images where the (latitude, longitude) of ISMN data point lies within the image bounds.
        iii. Perform a temporal join to retain images within a specified time bound in the past, where the bound depends upon the input source as specified in Table 5.
        iv. Pick the temporally closest image to the ISMN data point from the filtered images. If none are available, we discard this data point.
        v. Reproject the image to 10 m resolution and the corresponding UTM projection based on the UTM zone of the data point.
        vi. Crop the image to extract a 512x512 sized region centered around the (latitude, longitude) of the ISMN data point.
        vii. Normalize the image.
        viii. Pair this image with the data point.

Since each of the input sources have different revisit intervals that don't align perfectly with each other - eg.: Sentinel-1 (revisit of 6 days[2]) and Sentinel-2 (revisit of 5 days), we temporally anchor the data to Sentinel-1 (our primary high-res input) whilst setting temporal bounds for the other sources that allow us to acquire meaningful information. The next section describes the dataset created.

1) SENTINEL 1 ANCHORED DATASET

- For creating the dataset, we used Sentinel-1 data as the primary input. We chose to anchor upon Sentinel-1 since it is resilient to atmospheric conditions and is directly sensitive to soil moisture, unlike Sentinel-2 which primarily provides information for scene understanding. Data from other sources was used to enrich the Sentinel-1 inputs, so we allowed for a greater time slack there.
- The dataset is created using the Sentinel-1 anchored temporal bounds specified in Table 5. It is then split into train, validation, and test splits with a 60:20:20 ratio in an IID manner at a sensor level ie. 20% of the sensors are put under validation, 20% under test and the remaining are used for training. Data points belonging to a single sensor are always together.

---

[2] Sentinel-1 had a revisit period of 6 days before the Sentinel-1 B malfunction.

- The dataset comprises a total of 130,070 data points of which 85,591 (65.8%) data points are in the train split, 23,408 (18%) data points are in the validation split and 21,071 (16.2%) data points in the test split. Note that since all the sensors don't have the same number of datapoints, the 60:20:20 ratio is not strictly present across data points although it is present across sensors.
- Sample input imagery and labels are present in the Appendix.

| Source | Temporal bounds for Sentinel-1 anchored dataset |
|---|---|
| Sentinel-1 | 1 hr |
| Sentinel-2 | 14 days |
| SMAP | 3 days |
| GLDAS | 6 hrs |
| NASA DEM | 50 years (One time) |
| Soil Grids | 50 years (One time) |
| MODIS LAI | 8 days |

Table 5. Temporal bounds for each of the sources used during the join while creating the dataset. Note that in-situ readings are present on an hourly basis. The highlighted cell indicates the strictest bound (1hr) which ensures that each image in the source with the strictest bound is paired with a maximum of 1 in-situ sensor readings, as the sensor readings are never less than 1 hour apart. This ensures that we don't create duplicate data points in the dataset where the exact same set of imagery is paired with multiple in-situ readings.

2) DATASET STATISTICS

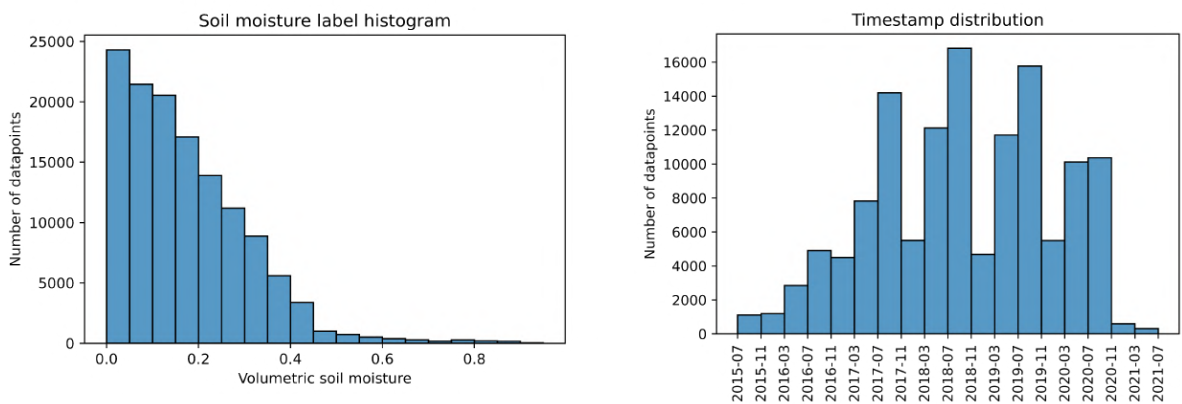

Fig. 1. *Left*: Volumetric soil moisture label distribution. *Right*: Timestamp distribution.

Volumetric soil moisture label distribution in Figure 1 shows that the labels are skewed towards the lower end, this can be attributed to the fact that most soils are not usually saturated. Additionally, cloud filtering on Sentinel-2 ends up dropping data points over cloudy intervals which are correlated to precipitation and hence higher soil moisture readings. This could induce a systemic bias in the dataset where the mean of the soil moisture labels is shifted towards the lower end. The timestamp distribution in the same figure shows that there's a fair spread of data across time, however, some seasons (seasonality considering the northern hemisphere since most of our data is from the northern hemisphere) have fewer data points noticeable as spiky dips, again due to cloud filtering on Sentinel-2.

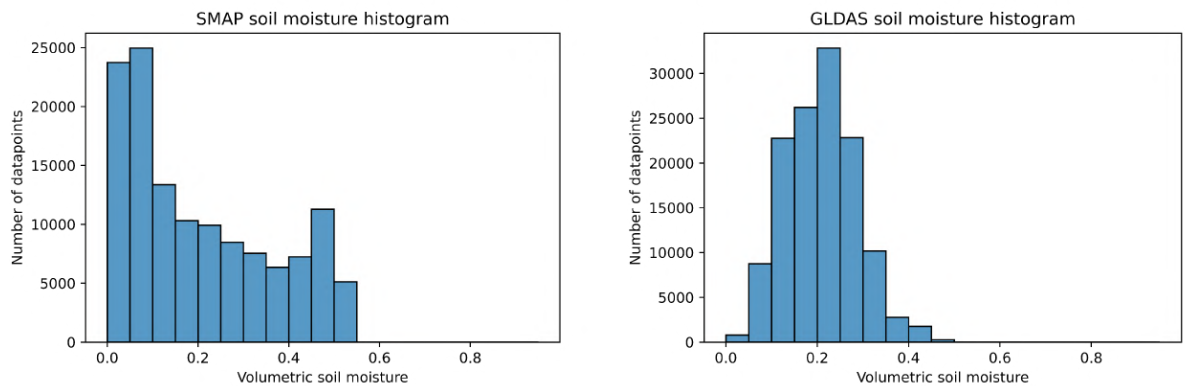

Fig. 2. Distribution of coarse soil moisture products. *Left*: SMAP volumetric soil moisture. *Right*: GLDAS volumetric soil moisture.

Distribution of coarse soil moisture products in Figure 2 shows that the SMAP soil moisture histogram resembles the label histogram in Figure 1 more closely when compared to the GLDAS histogram, likely because GLDAS estimates soil moisture at a depth of 10cm.

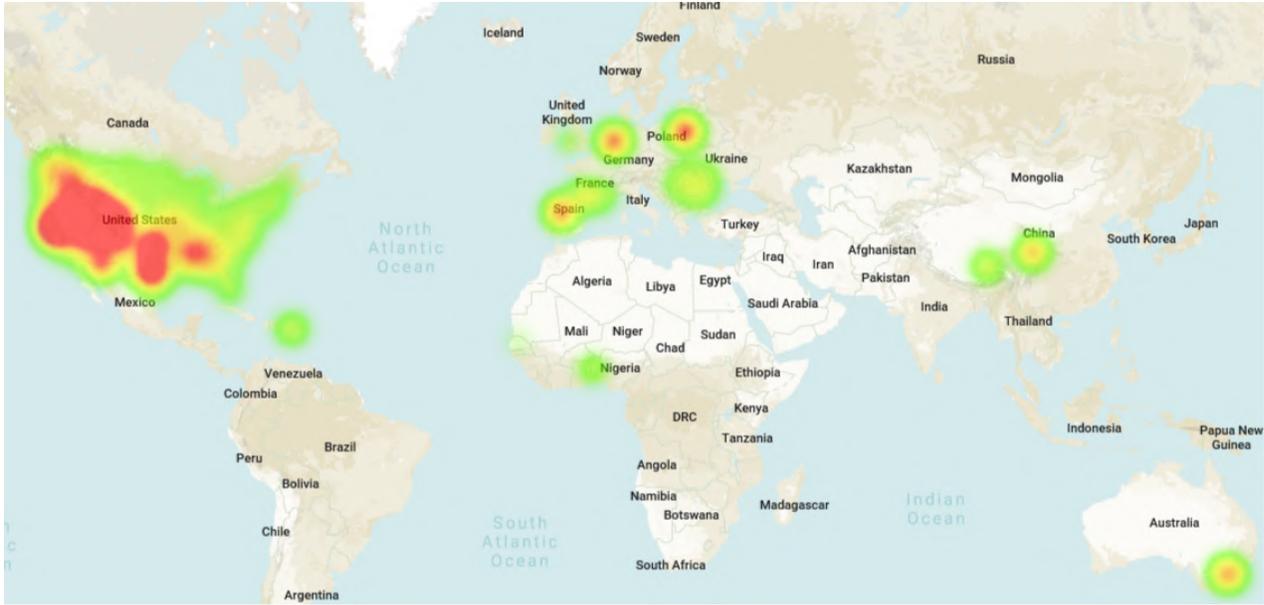

Fig. 3. A heatmap of in-situ sensor locations present in the dataset.

Figure 3 shows a heatmap of sensor locations present in the dataset. Around 83% of the sensors are in the US, 11% in Europe and the remaining around the globe. Note that this shows the distribution of in-situ sensors and not the actual data points so each sensor is represented only once.

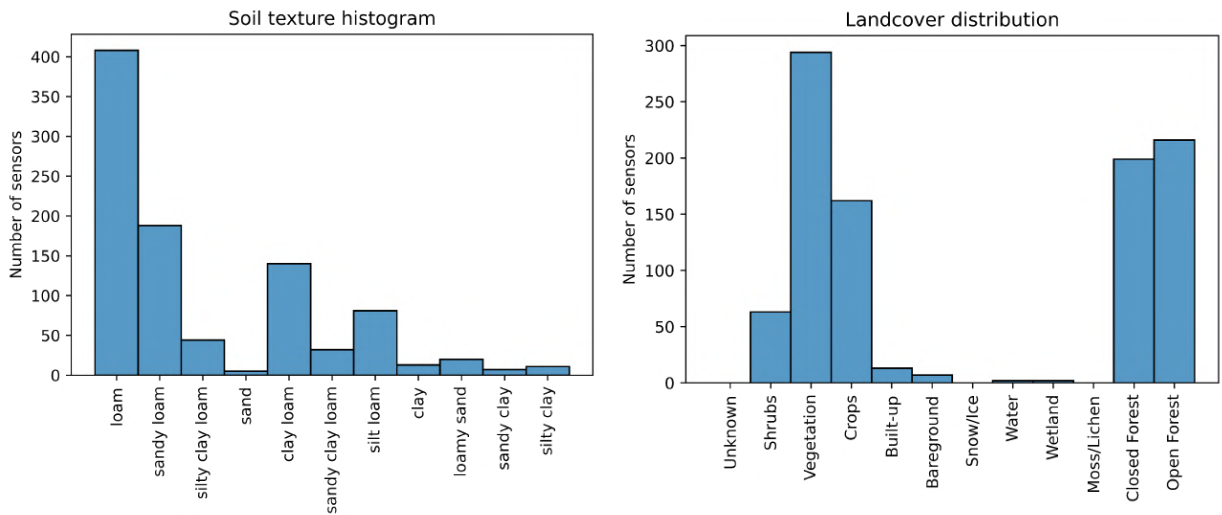

Fig. 4. *Left*: USDA based soil texture distribution. *Right*: Land cover distribution derived from the Copernicus Land Cover Map across the sensors present in the dataset.

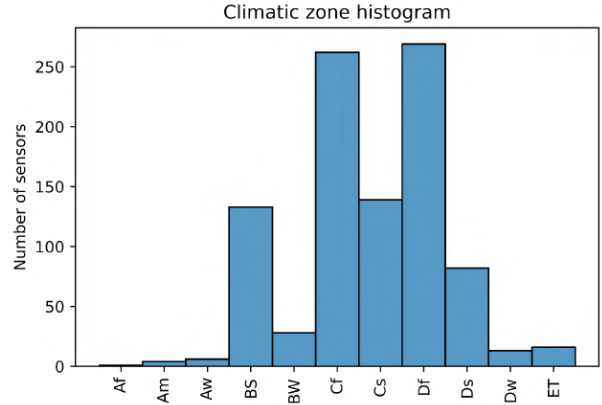

Fig. 5. Köppen climate zone distribution across the sensors present in the dataset. We use the first two identifiers of the Köppen classification only in order to group similar climatic zones together.

The soil texture distribution in Figure 4 shows the distribution of our sensors based on USDA soil texture classification (Groenendyk et al. 2015) where the input fractions of silt, sand, and clay were obtained from SoilGrids. Land cover distribution in Figure 4 shows that the majority of sensors are spread across vegetation, croplands and forests. This is a fairly broad distribution and allows us to see how our model performs across these different land cover types. We obtain land cover classes for each sensor from the Copernicus Land Cover Map (Buchhorn et al. 2020) on Earth Engine. Figure 5 shows the distribution of our sensors across the different climate zones (Kottek et al. 2006).

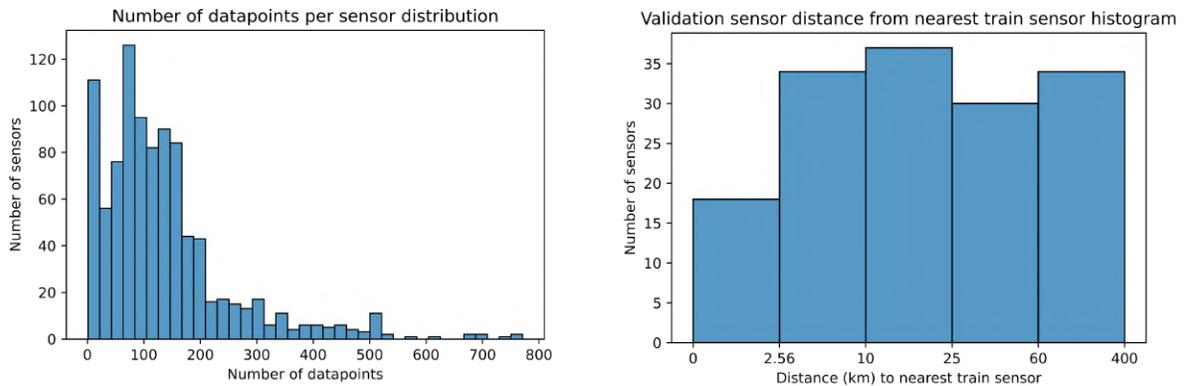

Fig. 6. *Left*: The distribution of the number of data points per sensor. *Right*: The distribution of the distance from each validation sensor to the nearest train sensor. Note that the bin width is constant and not to scale.

The distribution of the number of data points per sensor in Figure 6 shows that we have ~100-200 data points per sensor on average. This ensures that the data spans multiple seasons and years for a majority of the sensors. For each validation sensor that we evaluate on, we compute the geodesic distance to the nearest train sensor and plot these in a histogram in Figure 6. Only a small fraction of validation sensors (~12% of all validation sensors) are within the 2.56 km range (distance spanned by the high resolution sources) - mostly from the sensor grids we've used. We perform this analysis to ensure that the majority of validation sensors (~70%) are far away (>10 km) from their nearest train sensors to avoid any possible leakage of information and ensure we have a robust validation set.

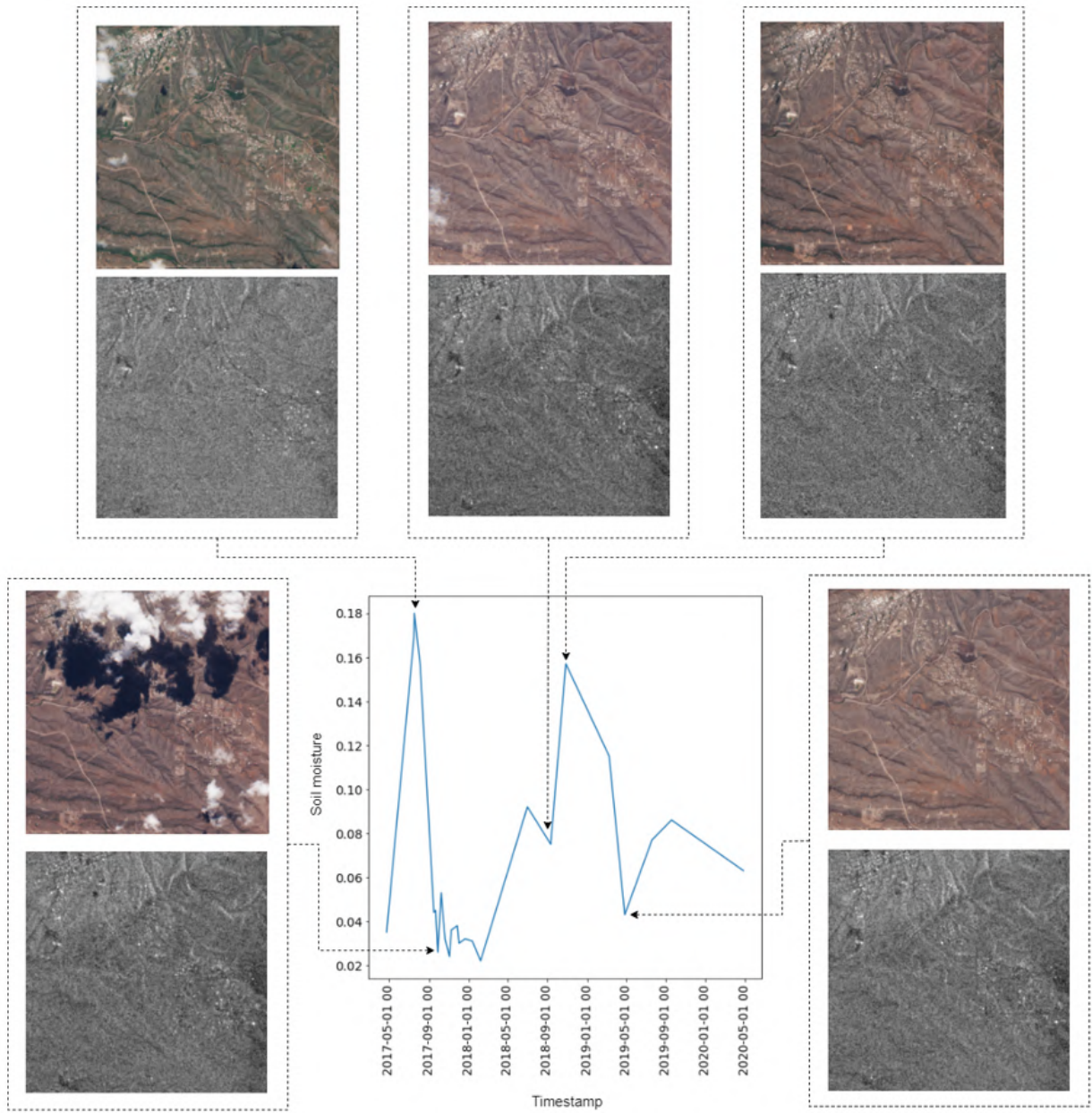

Fig. 7. Variation of soil moisture at a randomly selected sensor across time and corresponding input Sentinel-2 RGB and Sentinel-1 VV imagery.

Figure 7 captures the temporal variation of soil moisture at a single sensor. We observe that Sentinel-2 inputs can be quite visually indicative of the vegetation growth and dryness of a region which correlate with soil moisture.

## 3. Methods

The problem of soil moisture estimation is framed as an image based regression task (Babu, Zhao, and Li 2016; Fu et al. 2018; Rogez, Weinzaepfel, and Schmid 2017) where remote sensed sources and geophysical variables are used as inputs. These input sources provide the spatial and spectral covariates to estimate surface level soil moisture. We employ various deep learning techniques that have proven to work well for image based regression. These techniques allow our models to be site-agnostic, which results in better generalization compared to site-specific/calibrated models, providing the capability to scale globally.

*a. Model*

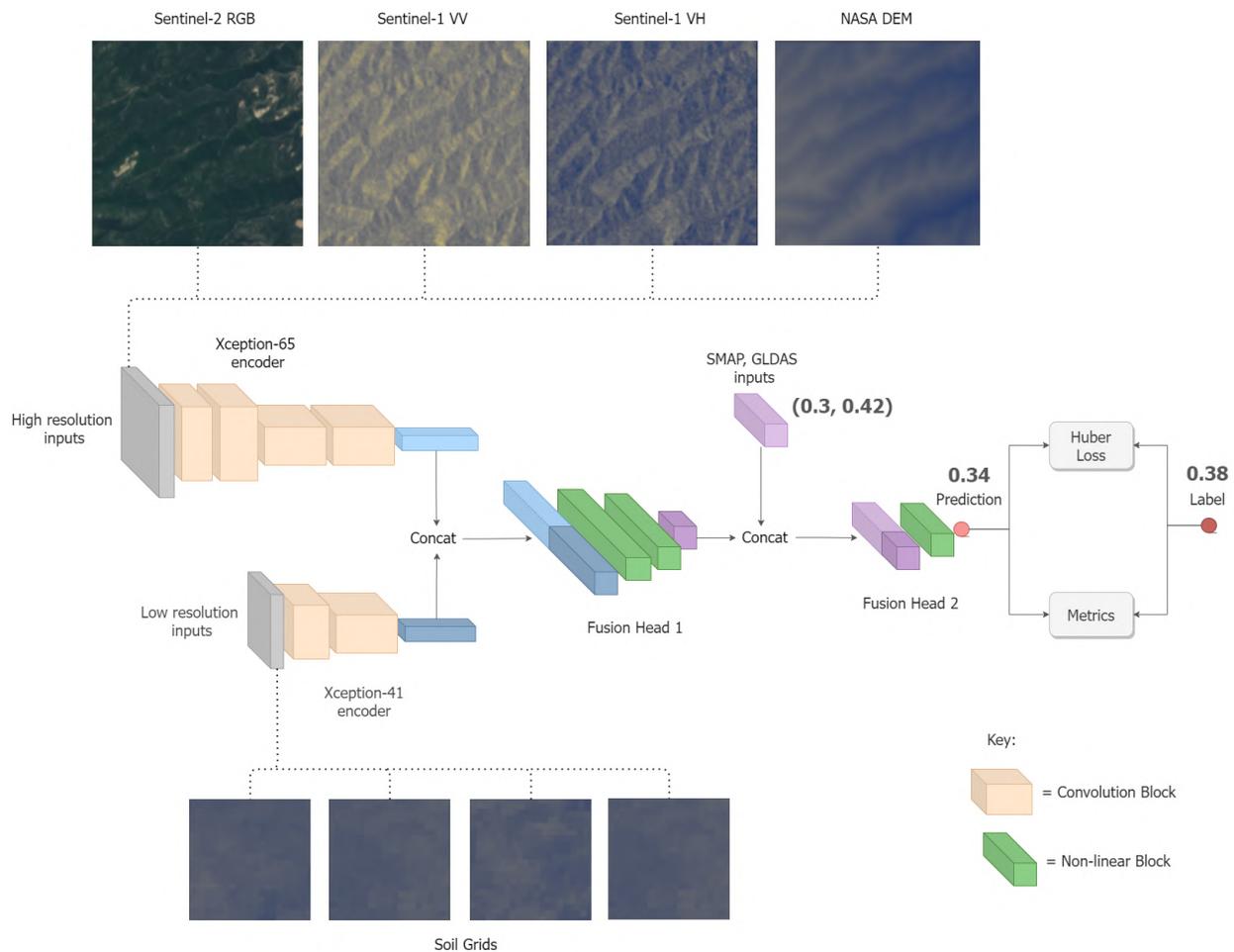

Fig. 8. Our model architecture for the task of soil moisture estimation.

Starting with the two sets of inputs (high-resolution and low-resolution), we encode each set of the inputs with Xception based feature encoders generating a single embedding (a compact feature representation) for each set of inputs. The hypothesis is that embeddings generated from the high resolution inputs capture a combination of the scene embedding (Main-Knorn et al. 2017; Raiyani et al. 2021) from Sentinel-2 and DEM sources and a potential rough soil moisture estimate (Paloscia et al. 2013) from the Sentinel-1 and Sentinel-2 sources. Likewise, the embeddings generated from the low res inputs would capture soil properties such as the water holding capacity, soil type, density and texture. These embeddings are then fused together by first concatenating both embeddings and then passing the concatenated vector through fusion head 1, which is a stack of [Dropout (Srivastava et al. 2014), Fully Connected (FC), Batchnorm (Ioffe and Szegedy 2015), Activation] layers followed by a [Dropout, FC] block at the end, described in detail in Section 3a2, resulting in a compact representation of features relevant for soil moisture.

We then fuse the coarse soil moisture inputs (SMAP and GLDAS SM estimates) with the output of the first fusion head, by passing the concatenated vector through another set of fully connected layers, called fusion head 2. This produces the final soil moisture estimate that is passed to the loss function to compare with training labels.

We also considered early-fusion approaches where the low resolution inputs are concatenated with the high-resolution inputs. However, these require the low resolution inputs to be scaled to the size of the high-resolution inputs which involves a large amount of duplication and wasted compute.

1) INPUT ENCODERS

We use the Xception (Chollet 2017) encoder for the feature encoders. We have also experimented with ResNet (He et al. 2016) and MobileNet-V2 (Sandler et al. 2018) encoders – empirically Xception performed the best out of these choices of encoders.

In particular, we use the Xception-65 encoder to encode high-resolution imagery and an Xception-41 to encode low resolution imagery. Low resolution imagery has a much smaller amount of input pixels which contain more semantically meaningful features and hence a smaller encoder with a fractional depth multiplier (Chollet 2017) works well and allows us to utilize model capacity better - faster training and fewer parameters.

The input to the high-res encoder consists of a 256 x 256 pixel image centered at the location of the sensor. High res imagery is sampled at a 10m resolution resulting in a 2.56 x 2.56 km region being used as the input. The low-res encoder uses a 16 x 16 input size where low res imagery (at 250m original resolution) is bilinearly resampled at 160m to ensure that we cover the same 2.56 x 2.56 km region. The large input region allows the model to understand the context around the center of the image to estimate the soil moisture at the center. The impact of varying the region size on the model performance is explored in the experiments section.

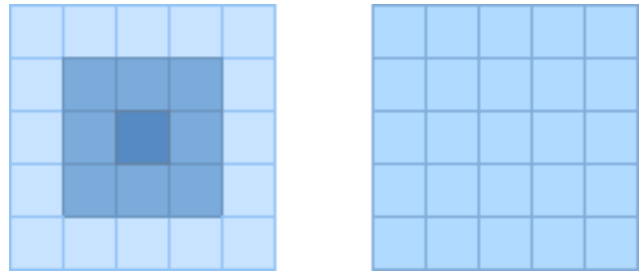

Fig. 9. A representative comparison of the features generated by the Xception encoder. *Left*: Center weighted pooling. *Right*: The standard average pooling Darker colors indicate a higher weight placed on a particular pixel. The sum of weights in both cases equals 1.

We apply a center weighted global pooling as shown in Figure 9 on the encoder features to ensure that the embeddings generated at the center are given the most importance since we are estimating the soil moisture at the center of the image.

2) FUSION HEAD

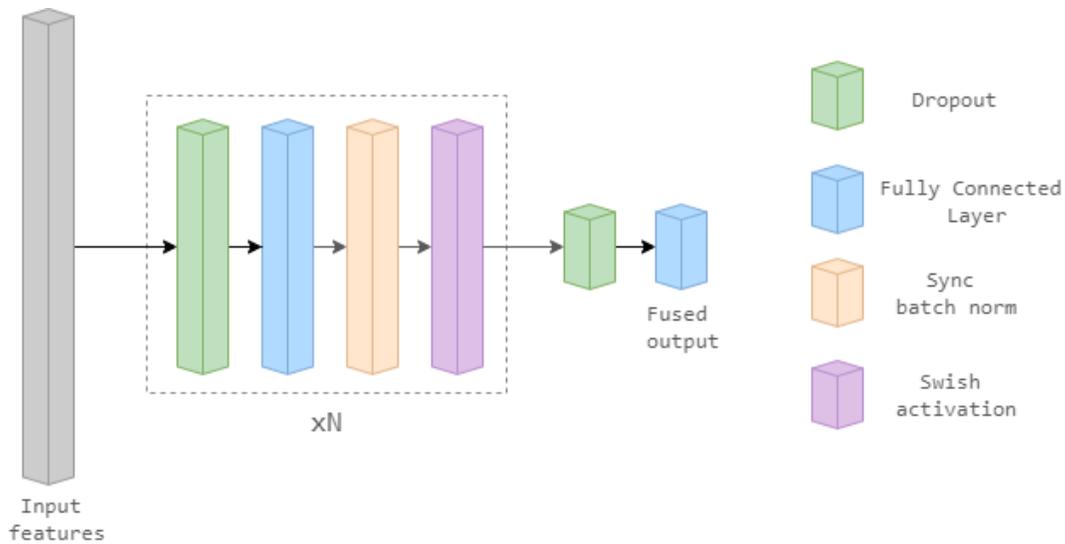

Fig. 10. Our fusion head architecture consists of fully connected, batch norm, activation and dropout layers. The dotted region shows one block of the fusion head which is repeated N times.

The fusion head consists of stacked blocks of [Dropout, Fully Connected (FC), Batchnorm, Activation] with an additional set of dropout and FC layers towards the end without an activation or batch normalization since we don't want to restrict our predicted output range. This head allows for nonlinear fusion of concatenated input features. Each fusion head can be configured by specifying the number of output channels per block and the number of blocks present.

We use two fusion heads in our model. The first one (with 3 blocks and output channels per block [128, 64, 1]) fuses information from the two input encoders encoding high-resolution and low resolution imagery respectively. The second one (with 2 blocks and output channels per block [8, 1]) fuses information from the coarse soil moisture inputs with the fused embedding generated by the first fusion head producing the final soil moisture estimate. Additional hyperparameter details are presented in Table 7.

## b. Training

We train our model using Tensorflow in a synchronized distributed TPU (v1) (Cheng et al. 2017) setting where each model is trained on a 2x2 pod (8 TPU cores). We use Stochastic Gradient Descent (SGD) with momentum (Rumelhart, Hinton, and Williams 1986) as the optimizer. Training takes around 8-10 hrs on average to complete. The learning rate transitions from the initial value to 0 via a polynomial decay schedule with a power of 0.9 across each minibatch during training. Additional hyperparameter details are presented in Table 7.

In addition to the distributed training strategy, we also utilize the Tensorflow data service (Murray et al. 2021) allowing us to distribute dataset pre-processing across a cluster of CPUs.

1) Loss

We use the Huber loss (Huber 1992) to train our models. The Huber loss is a combination of mean squared error (MSE) and mean absolute error (MAE) which allows our model to efficiently optimize inliers whilst paying lesser attention to outliers, delta specifying the threshold on determining what constitutes as an outlier.

$$loss = \begin{cases} \frac{1}{2} * (x - y)^2 & if\ (|x - y| \leq \delta) \\ \delta * |x - y| - \frac{1}{2} * \delta^2 & otherwise \end{cases}$$

(Huber, 1992)

We tried using MSE and MAE loss functions, but the Huber loss performed better empirically.

2) REGULARIZATION

Although the dataset consists of a large number of data points, it contains only 958 distinct locations since the number of sensors is limited. Due to this, our models could overfit to the small set of train locations. Regularization helps prevent this and improve generalization. We employ several regularization techniques to reduce overfitting such as Weight Decay (Loshchilov and Hutter 2019), Dropout (Srivastava et al. 2014) and Drop path (Huang et al. 2016)

*(i) Augmentations*

We also apply various augmentations on our inputs as an additional form of regularization (Shorten and Khoshgoftaar 2019). We only use center preserving augmentations since we are estimating the soil moisture at the center of the image (see discussion about center-weighted pooling in Section 3a1). We don't use any color transform augmentations since our remote sensing inputs have a much larger spectral range (i.e., not just RGB imagery) and we do not want to lose the signal present in the absolute value of pixels.

| Augmentation | Probability | Extent |
| --- | --- | --- |
| Rotate | 0.8 | +- 180 degrees |
| Flip Horizontal | 0.5 | None |
| Flip Vertical | 0.5 | None |
| Shear X | 0.3 | +-9 degrees |
| Shear Y | 0.3 | +- 9 degrees |

Table 6. Details on the exact set of augmentations used. The probability here indicates the probability with which a specific augmentation is applied on an input.

For every datapoint, each of the augmentations are applied based on their probabilities. This results in a large number of combinations of these augmentations allowing us to produce a wide variety of augmented images, similar to the augmentation policies described in autoaugment (Cubuk et al. 2019).

*c. Evaluation*

We evaluate our models both quantitatively and qualitatively. As a part of the qualitative evaluation, we plot multiple time series of model estimates and labels to visually assess temporal variation at randomly selected sensors. This is useful to identify whether any biases are present at different sensor locations, and to visually assess how biases might differ based on properties of the location such as the soil type, land cover etc. For quantitative evaluation, we look at a set of metrics on the best model checkpoint during training as described below.

1) METRICS

To measure the performance of our models, we compute various metrics for each sensor and then average them over all our sensors where at least 13 data points are present at each sensor.

$$\text{final\_metric} = \frac{\sum_{s=1}^{M} metric_s}{M}$$

Where,
**M** = number of sensors having >= 13 data points
**metric$_s$** = per sensor metric

The metrics that we report are: (i) ubRMSE, (ii) RMSE, and (iii) correlation (R), which are described as follows:

Notation used in the metrics below,
**x$_i$** = Model estimate
**y$_i$** = Label
**x̄** = Mean model estimate
**ȳ** = Mean of labels at a particular sensor location
**N** = Number of data points

*(i) ubRMSE*

A common metric used to measure the performance of soil moisture products is the unbiased RMSE (ubRMSE). This metric computes the RMSE after subtracting the mean from both labels and model estimates, respectively. It is often the case that soil moisture estimates have large climatological differences with in-situ sensors, however, we are often concerned with per-location dynamics instead of absolute values. The ubRMSE is a common soil moisture performance metric that addresses these points.

$$\text{ubRMSE} = \sqrt{\frac{\sum_{i=1}^{N}\left(\left(x_i - \bar{x}\right) - \left(y_i - \bar{y}\right)\right)^2}{N}}$$

*(ii) RMSE*

Reporting RMSE in addition to ubRMSE allows insight into per-sensor bias.

$$RMSE = \sqrt{\frac{\sum_{i=1}^{N}\left(x_i - y_i\right)^2}{N}}$$

*(iii) Correlation*

RMSE and ubRMSE are unbounded metrics, and therefore difficult to contextualise outside of inter-model comparisons. We therefore also report the standard Pearson product-moment correlation coefficient:

$$r = \frac{\sum_{i=1}^{N}(x_i - \bar{x})(y_i - \bar{y})}{\sqrt{\sum_{i=1}^{N}(x_i - \bar{x})^2 \sum_{i=1}^{N}(y_i - \bar{y})^2}}$$

## *d. Hyperparameter tuning*

| Hyper-parameter name | Value/Range | | | |
|---|---|---|---|---|
| | **High-res encoder** | **Low-res encoder** | **Fusion head 1** | **Fusion head 2** |
| **Model** | | | | |
| Encoder architecture | Xception-65 | Xception-41 | - | - |
| Input crop size | 256 selected from (512, 256, 128, 64) | 16 | - | - |
| Output stride | 32 | 16 | - | - |
| Drop-path keep probability | 0.6 selected from (0.8, 0.6, 0.4, 0.2) | | - | - |
| Fusion head shape | - | - | (128, 64, 1) | (8, 1) |
| Fusion head dropout keep probability | - | - | (0.5, 0.5, 0.6) | (0.9, 0.9) |
| Fusion head activation | - | - | Swish selected from (Swish, ReLU, Leaky ReLU) | |
| **Training** | | | | |
| Batch size | 64 | | | |
| Optimizer | Momentum (m=0.9) | | | |
| Number train steps | 600000 | | | |

| Learning rate | Sweep across (0.1, 0.05, 0.03) |
| --- | --- |
| Weight decay | Sweep across (0.0001, 0.00005, 0.00003) |
| Huber loss delta | 0.4 selected from (0, 0.2, 0.4, 0.6, 0.8, 1.0) |

Table 7. Training hyper-parameters used for all of the experiments.

For each of our experiments, we run a grid sweep on a set of initial learning rates and the weight decays as specified in Table 7 and pick the model that performed best on the evaluation set.

We tried using ImageNet checkpoints to initialize the Xception encoders since ImageNet based initializations have shown to perform well in various transfer learning scenarios (Kornblith, Shlens, and Le 2019), but we did not see any significant improvement in performance for the task at hand. Hence, we just use random initialization for the weights of our models. We use a global batch size of 64, which is split across the 8 TPUs for training (data parallel distribution) with synchronized batch norm. We trained each model for a total of 600k steps to ensure training curves plateau (not reported).

*e. Model resolution*

Our models produce soil moisture estimates at a nominal resolution of 320m. This is because our encoders reduce the high resolution input (256 x 256 pixels @ 10m per pixel) by a factor of 32 to create a feature representation that is 8 x 8 pixels @ 320m per pixel. We use the term "nominal" because, although the model produces representations at a 320m resolution, nearby predictions are not completely independent and even when we stride out inputs at 320m to get estimates for adjacent locations, the inputs still overlap.

## 4. Experiments

We adopt the following model naming convention: the names of all the sources that are used as inputs to the model are concatenated to identify the model. Please refer to Figure 8 for details on how each of the inputs are used.

Hereafter, we refer to the Sentinel-1 + DEM + Sentinel-2 + SoilGrids + SMAP + GLDAS model as "our model" and will be the model used in experiments, results and other analyses unless specified otherwise.

*a. Overall model performance*

The following set of baselines and ablations are used to measure the performance of our model quantitatively. Results are reported on both the validation and the test set.

- **Baseline**:

- **SMAP + GLDAS NN** - The coarse soil moisture inputs passed through a fusion head to produce a soil moisture estimate.
- **Sentinel-1 + DEM + Sentinel-2 + SoilGrids + SMAP + GLDAS** - Our model as described in Figure 8.
- **Ablations**:
    - **Sentinel-1 + DEM** - A Sentinel-1 only model (DEM allows the model to factor in the terrain)
    - **Sentinel-1 + DEM + Sentinel-2** - A Sentinel-1 and Sentinel-2 combined model where both of the inputs are passed together to the high resolution encoder whose embeddings are then passed through a soil moisture head.
    - **Sentinel-1 + DEM + Sentinel-2 + SoilGrids** - Similar to the previous model, except that SoilGrids is also used.

## b. Spatial and Temporal Analysis

### 1) Spatial stratification of model performance

To understand model fidelity, multiple stratified analyses are performed. We look at the performance variation across land cover classes, soil texture types and climate zones. The Copernicus Global Land Cover map is used for the land cover class, SoilGrids for the soil texture type and the Köppen climate map for the climate zone.

To understand variation at a finer scale, quantitative analysis at an individual sensor level is also performed. This is done for each of the validation sensors present in the US (since a majority of the sensors are located in the US)

### 2) Time series analysis

To understand temporal variations in the performance of our model, we look at the time series of model estimates vs in-situ labels for a few randomly selected validation sensors. This helps understand the temporal stability of model estimates and visualize the bias present (if any).

### 3) Large scale spatial inference

Lastly, large scale inference is performed using our model where we move the model pixel by pixel and estimate the soil moisture at each location. This provides insight into spatially coherency of the model estimates and variation with respect to the inputs.

## c. Benchmarking

We pick some of the best existing global methods along with a few top regional models to compare and evaluate our model performance against. Information about the works we compare against is presented in Table 8.

| Work | Soil moisture label depth | Temporal range of in-situ data | Validation sensor networks | Validation information | Additional details |
|------|---------------------------|-------------------------------|----------------------------|------------------------|--------------------|
| (Y. Kerr et al. 2016) | 0-5 cm | 2010-2013 | AMMA-CATCH, SCAN, SNOTEL, USCRN | All sensors used for validation. | SMOS validation. Only provide median correlation metrics instead of mean across sensors in each of their networks. Still a useful comparison though since the median is often better calibrated and robust to outliers. |
| (Bi et al. 2016) | 0-5 cm | 2010-2012 for NAQU, 2008-2011 for MAQU | MAQU, NAQU | All sensors used for validation. In-situ data averaged in each LSM pixel. | Validation of 4 different Land Surface Models (LSMs) - VIC (Best in MAQU), Noah (GLDAS), Mosaic, CLM (Best in NAQU). We compare against the best for each network. |
| (Albergel et al. 2012) | 0-5 cm (Model estimates at 0-7 cm though) | 2008-2010 | AMMA-CATCH, SCAN, SMOSMANIA, REMEDHUS | All sensors used for validation. | ECMWF soil moisture validation. No ubRMSE metrics reported. However, the RMSE and bias are reported per sensor for some of the sensor networks, so we compute the ubRMSE as sqrt(RMSE$^2$ - bias$^2$) per sensor and then take an average to get the ubRMSE for the network. |
| (Wang et al. 2021) | 0-5 cm | 2017-2018 | FR_Aqui, RSMN, SCAN, SMOSMANIA, SNOTEL, USCRN, REMEDHUS | All sensors used for validation. Sensors within a single CCI/SMOPS pixel are averaged. All readings are averaged over | They study two products CCI and SMOPS, CCI performs better overall so we compare against that. Only correlation and RMSE metrics present, since bias metrics aren't provided, we can't compute the ubRMSE. |

| | | | | 24 hrs. | |
|---|---|---|---|---|---|
| (Fang et al. 2021) | 0-5 cm | 2018-2019 | SCAN, USCRN | All sensors used for validation. | They produce soil moisture estimates at 400m resolution using a statistical approach. |
| (Balenzano et al. 2021) | 0-5 cm | 2015-2018/ 2020 (for some sites) | LITTLEWASHITA, REMEDHUS, TXSON, YANCO | All sensors used for validation. | They obtain data from ISMN along with a couple of SMAP validation sites. They use temporal Sentinel-1, land cover and soil texture inputs in a physics based method. |
| (Dente, Su, and Wen 2012) | 0-5 cm | 2010 | MAQU, TWENTE | All sensors used for validation. In-situ data averaged in each SMOS pixel. | Validation of SMOS retrievals in MAQU and Twente. |
| (Collow et al. 2012) | 0-5 cm | 2010 | FORTCOBB, LITTLEWASHITA | All sensors used for validation. In-situ data averaged in each SMOS pixel. | Validation of SMOS retrievals in FortCobb and LittleWashita. |
| (Jing, Song, and Zhao 2018) | 0-5 cm | 2008-2011 | KYEAMBA, YANCO | All sensors used for validation. | They validate ECMWF based multilayer temporal LSMs. |
| (Karthikeyan and Mishra 2021) | 0-5 cm | 2015-2019 | FORTCOBB, LITTLEWASHITA, TXSON | Sensor level splits. | Build and use multiple region specific XGBoost models. Only provide median metrics for each of their networks. |
| (Abbaszadeh, Moradkhani, and Zhan 2019) | 0-5 cm | 2015 | SCAN, USCRN, LITTLEWASHITA | 80:20 train/val split over entire data (not at a sensor level). | They use a set of 12 random forest models where each model is used for a different soil texture type. |

Table 8. Information on the works we compare against.

Not all methods in Table 8 use the same kind of train/eval splits and acquisition time ranges for in-situ data, so it's not possible to make rigorous comparisons. However, the sensor networks used are the same in all comparisons and sampling is always performed over a large time range which provides a meaningful comparison. A lot of works also perform K-fold cross validation while reporting their results, but we don't do that on our end due to practical constraints - deep learning models take a long time to train. We instead do a rigorous train/validation/test split, where we only looked at the numbers on the test set exactly once (at the time of writing this paper), and after all model parameters were finalized on the validation split.

*d. Model exploration studies*

Ablation studies and sensitivity analyses are performed to see how important various model inputs/parameters are. All the studies here were performed on the validation set.

1) INPUT SIZE SENSITIVITY ANALYSIS

Each of the high-resolution sources correspond to 10x10m per pixel. The sensitivity study here uses the Sentinel-1 + Sentinel-2 + DEM model. We only use high resolution sources here since they are the most sensitive to a change in input size. This helps us understand the amount of context required by the model better and how it ties to performance.

2) INPUT FEATURE ABLATION STUDY

To identify which features are the most important for our models, we perform the following experiments. Starting with our model with all the input bands (Sentinel-1, DEM, Sentinel-2, SoilGrids and the coarse soil moisture inputs - SMAP, GLDAS) as the reference model, we remove the source for which we want to calculate the feature importance and then see how much of a drop in performance is observed.

# 5. Results

*a. Overall model performance*

|  | Validation | | | Test | | |
| --- | --- | --- | --- | --- | --- | --- |
| **Experiment** | ubRMSE | RMSE | Correlation | ubRMSE | RMSE | Correlation |
| **Baselines (Low-resolution)** | | | | | | |
| SMAP | 0.097 | 0.144 | 0.638 | 0.1 | 0.134 | 0.638 |
| GLDAS | 0.07 | 0.114 | 0.572 | 0.07 | 0.11 | 0.58 |
| SMAP + GLDAS NN | **0.061** | **0.102** | **0.663** | **0.063** | **0.099** | **0.668** |
| **Ours (High-resolution)** | | | | | | |
| Sentinel-1 + DEM | 0.073 | 0.099 | 0.474 | 0.075 | 0.109 | 0.437 |
| Sentinel-1 + DEM + Sentinel-2 | 0.067 | 0.094 | 0.587 | 0.069 | 0.099 | 0.56 |
| Sentinel-1 + DEM + Sentinel-2 + SoilGrids | **0.058 (+4.9%)** | **0.089 (+12.7%)** | **0.675 (+1.8%)** | **0.06 (+4.7%)** | **0.096 (+3%)** | 0.647 (-3.1%) |
| Sentinel-1 + DEM + Sentinel-2 + SoilGrids + SMAP + GLDAS | **0.054 (+11.5%)** | **0.085 (+16.7%)** | **0.727 (+9.7%)** | **0.055 (+12.7%)** | **0.088 (+11.1%)** | **0.729 (+9.1%)** |

Table 9. Validation and test results. A '+' in percentage change denotes an increase in correlation (decrease in ubRMSE) relative to the SMAP + GLDAS NN baseline. The validation split consists of 23,408 data points and the test split consists of 21,071 data points.

A quantitative evaluation of our models is presented in Table 9 along with comparisons to baselines. Our model performs significantly better than the baseline SMAP + GLDAS NN in all the metrics. The Sentinel-1 + DEM + Sentinel-2 + SoilGrids model (which does not use any coarse soil moisture inputs) performs slightly better than the baseline SMAP + GLDAS NN (except the test set correlation). These results together show that the high resolution sources/geophysical variables and coarse soil moisture sources provide complementary information and combining them gives us the best performance.

## b. Spatial and Temporal Analysis

1) SPATIAL STRATIFICATION OF MODEL PERFORMANCE

| | Validation | | |
|---|---|---|---|
| **Land cover** | **ubRMSE** | **RMSE** | **Correlation** |
| Agriculture | 0.055 | 0.084 | 0.729 |
| Closed forest | 0.057 | 0.082 | 0.726 |
| Open forest | 0.055 | 0.088 | 0.745 |
| Vegetation | 0.055 | 0.089 | 0.71 |
| Shrubs | 0.034 | 0.068 | 0.775 |
| Bareground* | 0.027 | 0.029 | 0.725 |
| Built-up* | - | - | - |
| Wetland* | - | - | - |

Table 10. Validation results for our model stratified by land cover type. An asterisk (*) indicates that there are < 5 sensors available for the specific land cover class in the test data.

| | Validation | | |
|---|---|---|---|
| **Soil texture type** | **ubRMSE** | **RMSE** | **Correlation** |
| Loam | 0.052 | 0.085 | 0.72 |
| Sandy loam | 0.055 | 0.083 | 0.78 |
| Silty clay loam | 0.056 | 0.079 | 0.57 |
| Clay loam | 0.05 | 0.088 | 0.7 |
| Sandy clay loam | 0.032 | 0.074 | 0.78 |
| Silt loam | 0.056 | 0.079 | 0.78 |
| Sand* | 0.032 | 0.032 | 0.15 |
| Clay* | 0.052 | 0.06 | 0.67 |
| Loamy sand* | 0.108 | 0.117 | 0.77 |
| Sandy clay* | 0.115 | 0.134 | 0.8 |
| Silty clay* | - | - | - |

Table 11. Validation results for our model stratified by soil texture type. An asterisk (*) indicates that there are < 5 sensors available for the specific soil texture class in the test data.

| | Validation | | |
|---|---|---|---|
| **Climatic zone** | **ubRMSE** | **RMSE** | **Correlation** |
| Semi arid (BS) | 0.063 | 0.086 | 0.68 |
| Dry desert (BW) | 0.037 | 0.043 | 0.76 |
| Humid subtropical (Cf) | 0.05 | 0.077 | 0.73 |
| Summer Mediterranean (Cs) | 0.049 | 0.095 | 0.77 |
| Humid continental (Df) | 0.059 | 0.094 | 0.75 |
| Mediterranean-influenced (Ds) | 0.057 | 0.069 | 0.8 |
| Monsoon influenced (Dw)* | 0.059 | 0.094 | 0.45 |
| Tundra (ET)* | 0.032 | 0.078 | 0.9 |
| Tropical rainforest (Af)* | - | - | - |
| Tropical monsoon (Am)* | - | - | - |
| Tropical wet and dry (Aw)* | 0.033 | 0.044 | 0.93 |

Table 12. Validation results for our model stratified by the climatic zone. An asterisk (*) indicates that there are < 5 sensors available for the specific climate zone in the validation data.

Our model performs well across different kinds of land cover classes as shown in Table 10, soil texture types as shown in Table 11 and climatic zones as shown in Table 12. Metrics on classes containing less than 5 sensors should be disregarded since the sample size is too small to get a reliable aggregate. In all of these stratifications, the model performs fairly consistently (correlation within +-0.15 of the overall correlation metric of 0.727) across all the classes showing adaptability and robustness of the model.

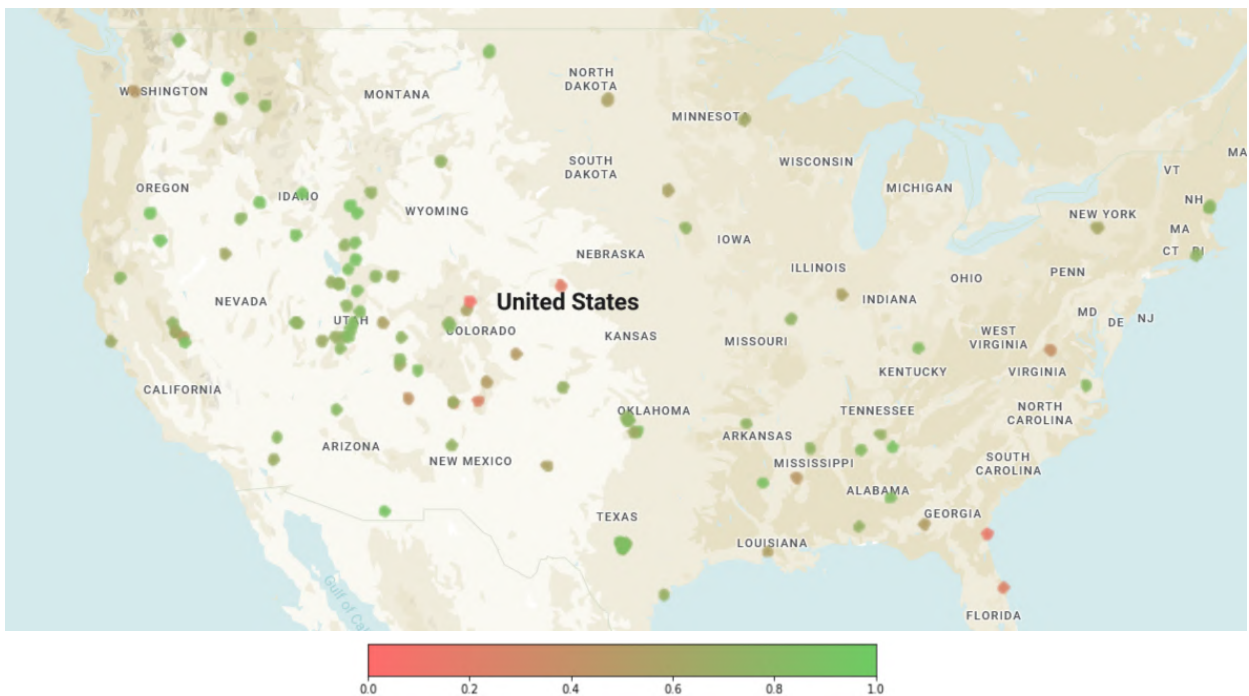

Fig. 11. Correlation between model estimates and ground truth labels at each validation sensor location in the US. Red-Green circles indicate in-situ sensor locations at which we report the correlation.

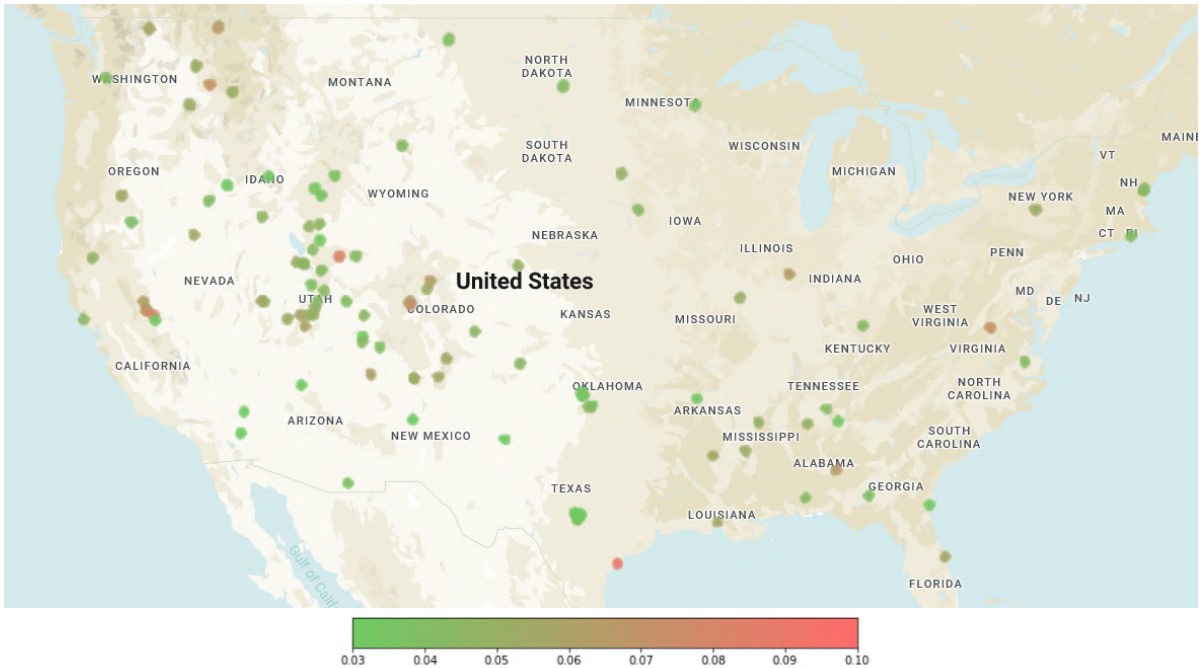

Fig. 12. ubRMSE between model estimates and ground truth labels at each validation sensor location in the US. Red-Green circles indicate in-situ sensor locations at which we report the ubRMSE.

Figure 11 shows the correlation at a sensor level for each of our validation sensors across the US. The model performs poorly at a few locations in Colorado and along the coastline of Florida. Sensors along the coastlines show a slight drop in performance overall. Figure 12 shows a similar map but for the ubRMSE metric.

## 2) TIME SERIES ANALYSIS

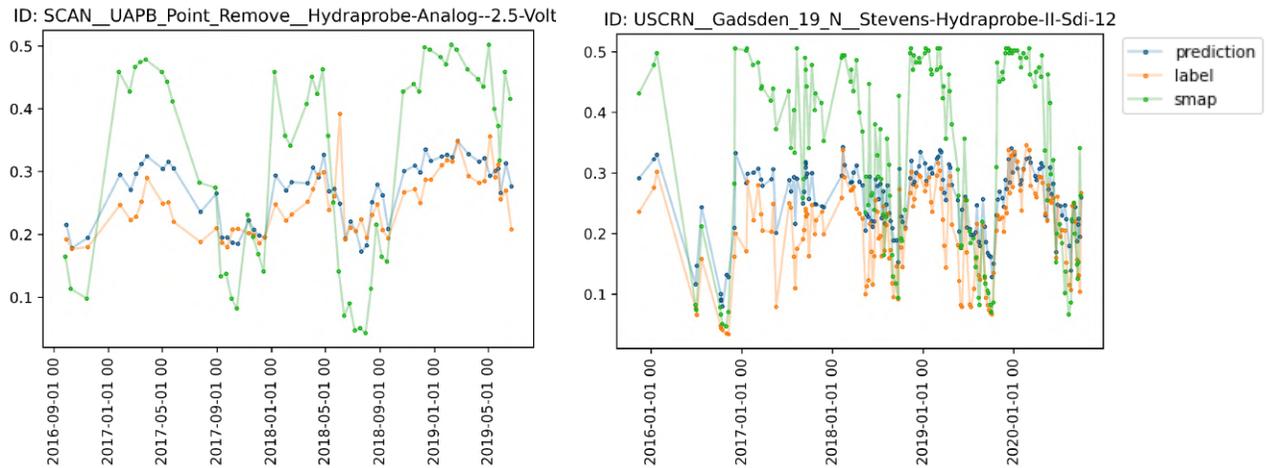

Fig. 13. Sample time-series plots showing how the estimates of the model vary over time at a few randomly picked validation sensor locations compared to the labels/SMAP estimates. The sensor corresponding to the figure on the left has the following properties: land cover - agriculture, soil type - silt loam, climate zone - Cfa. The sensor on the right has the same properties as the one on the left except the soil type which is loam.

Figure 13 shows time series plots for a few randomly selected sensors. In a majority of cases, our model estimates follow the variation in the sensors well. They often have a small amount of bias based on the location but the overall trend is captured well by the model. This reflects what we see in our quantitative metrics - correlation and ubRMSE.

## 3) LARGE SCALE SPATIAL INFERENCE

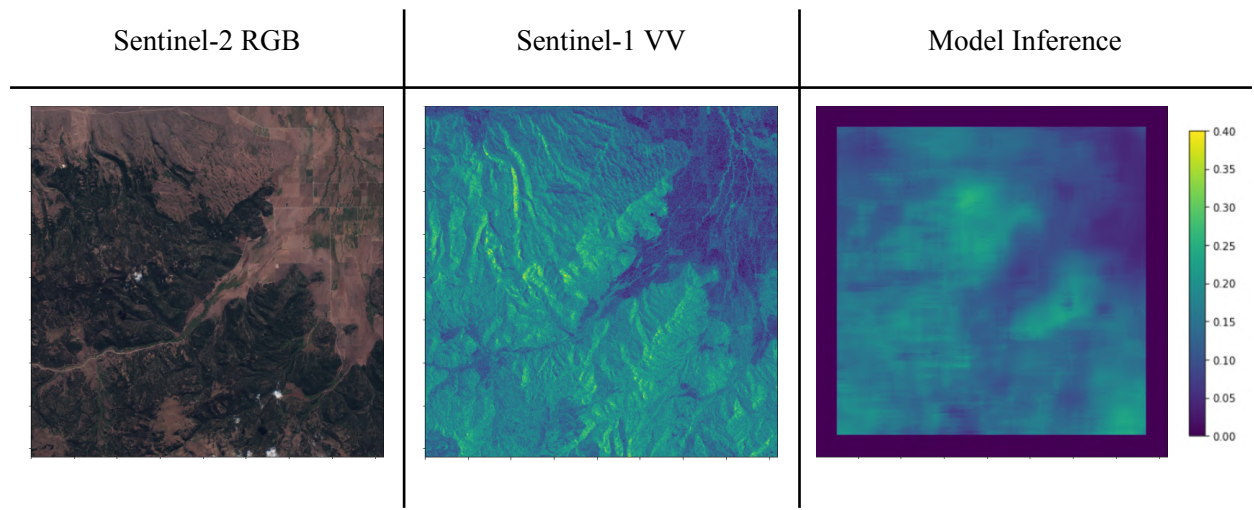

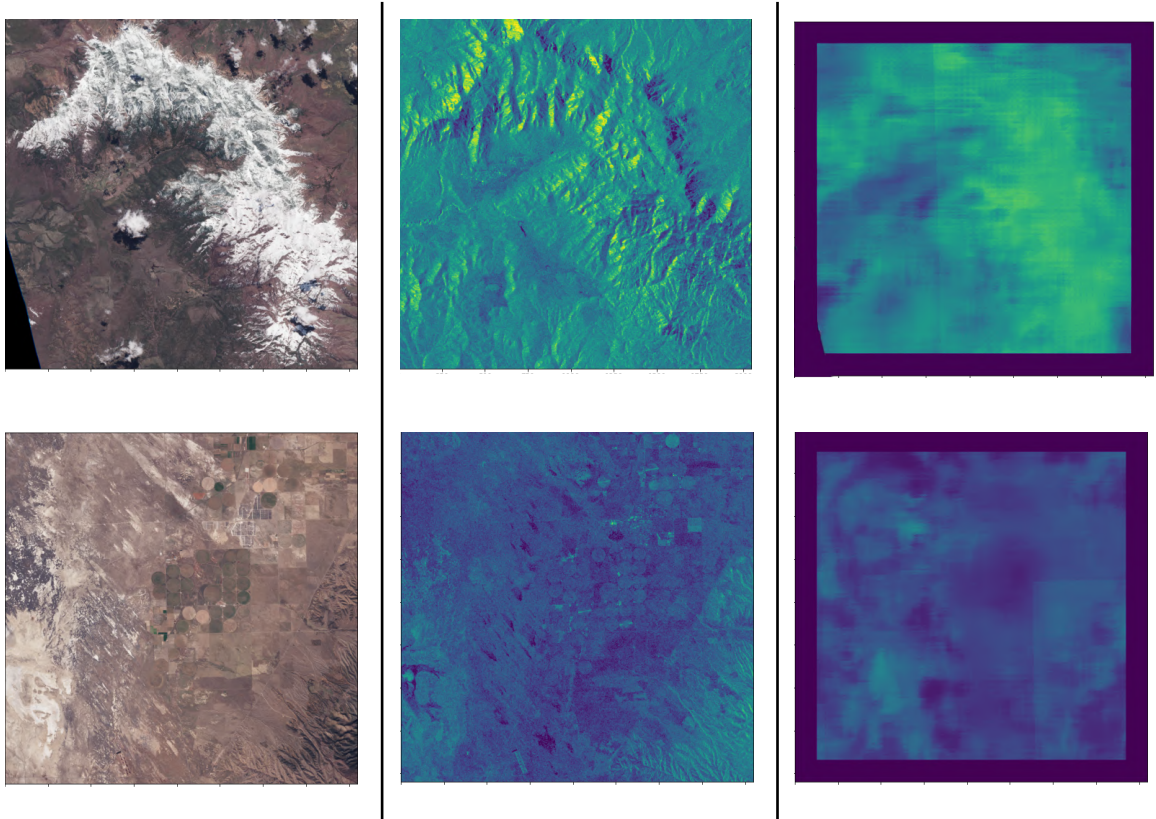

Fig. 14. Sample inference results on our model on ~20x20km regions selected at random. We notice a fair amount of variation at a sub-SMAP pixel (9x9km) level as well. The local variation in the estimates can be explained to a large extent by looking at the input imagery. These results along with the metrics we observe in Table 9 show that the model produces accurate high resolution estimates of soil moisture.

| Sentinel-2 RGB | Sentinel-1 VV | Model Inference |
| --- | --- | --- |

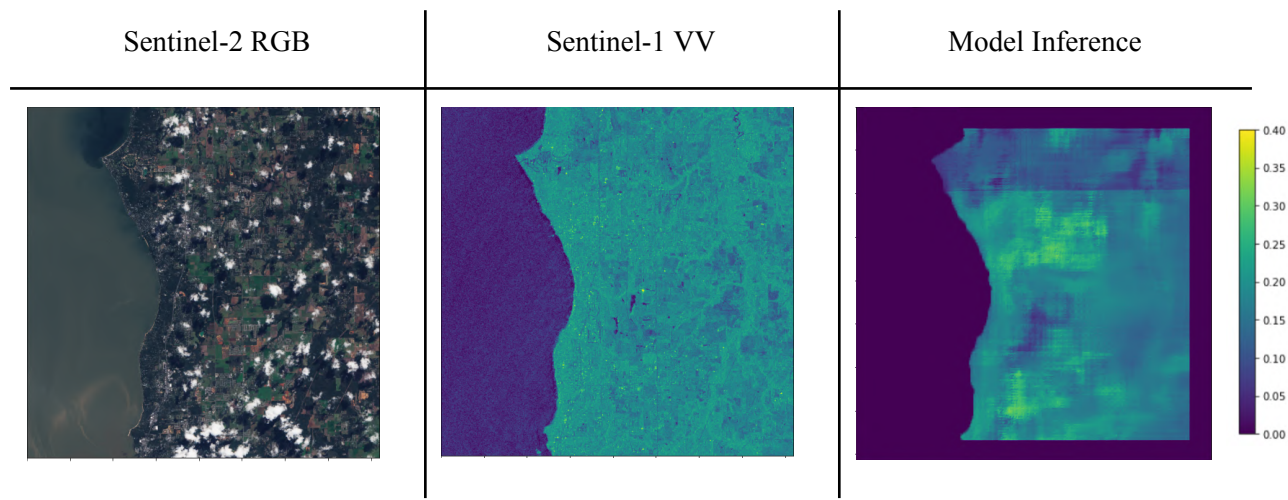

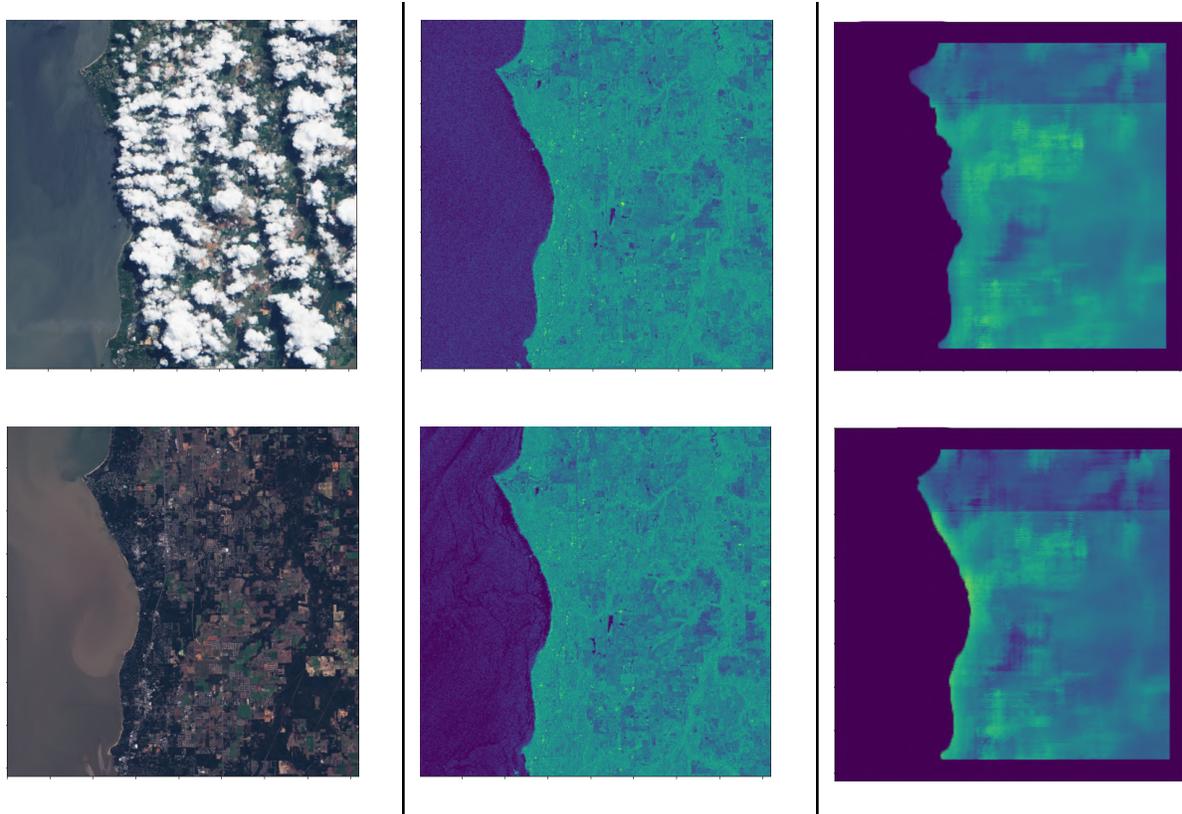

Fig. 15. Model inference results across time (each row separated by a year) on a single ~20x20km region selected at random. Notice that although the actual estimated values of soil moisture change, the patterns visible remain fairly consistent which shows the reliability of the model. At times however, tiling artifacts at SMAP boundaries appear when there is a large variation in adjacent SMAP tiles.

*c. Benchmarking*

| Network | Correlation | | | | | | | | | | | | | |
|---|---|---|---|---|---|---|---|---|---|---|---|---|---|---|
| | Ours | Kerr et al[3] | Bi et al[4]* | Albergel et al[5]* | Wang et al[6] | Fang et al[7] | Balenzano et al[8] | Dente et al[9]* | Collow et al[10]* | Jing et al[11]* | Karthikeyan et al[12]** | Abbaszadeh et al[13]*** | SMAP | GLDAS |

[3] (Y. Kerr et al. 2016) (SMOS)
[4] (Bi et al. 2016)
[5] (Albergel et al. 2012) (ECMWF)
[6] (Wang et al. 2021)
[7] (Fang et al. 2021)
[8] (Balenzano et al. 2021)
[9] (Dente, Su, and Wen 2012)
[10] (Collow et al. 2012)
[11] (Jing, Song, and Zhao 2018)
[12] (Karthikeyan and Mishra 2021)
[13] (Abbaszadeh, Moradkhani, and Zhan 2019)

| Network | Ours | Bi et al[14]* | Albergel al[15]* | Fang et al[16] | Balenzano et | Collow et | Jing et | Karthikeyan | Abbaszadeh | SMAP | GLDAS |
|---|---|---|---|---|---|---|---|---|---|---|---|
| AMMA-CATCH | **0.93** | 0.77 | | 0.53 | | | | | | 0.876 | 0.872 |
| BIEBRZA_S-1 | **0.902** | | | | | | | | | 0.505 | 0.461 |
| FR_Aqui | **0.848** | | | 0.78 | | | | | | 0.836 | 0.801 |
| MAQU | 0.450 | | 0.623 | | | | **0.63** | | | 0.509 | 0.54 |
| NAQU | **0.898** | | 0.826 | | | | | | | 0.806 | 0.802 |
| RSMN | 0.428 | | | **0.62** | | | | | | 0.339 | 0.381 |
| SCAN | **0.665** | 0.6 | | 0.63 | 0.55 | 0.656 | | | 0.65 | 0.654 | 0.541 |
| SMOSMANIA | **0.88** | | | 0.8 | 0.67 | | | | | 0.868 | 0.757 |
| SNOTEL | **0.736** | 0.48 | | | 0.48 | | | | | 0.609 | 0.485 |
| USCRN | 0.674 | **0.7** | | | 0.58 | 0.688 | | | **0.7** | 0.597 | 0.628 |
| FORTCOBB | 0.767 | | | | | | 0.72 | **0.77** | | 0.615 | 0.65 |
| KYEAMBA | **0.913** | | | | | | | 0.84 | | 0.899 | 0.883 |
| LITTLEWASHITA | 0.781 | | | | | 0.28 | 0.76 | 0.77 | **0.82** | 0.66 | 0.578 |
| REMEDHUS | **0.786** | | | 0.77 | 0.65 | 0.2 | | | | 0.737 | 0.735 |
| TWENTE | **0.799** | | | | | | 0.68 | | | 0.706 | 0.578 |
| TXSON | 0.821 | | | | | 0.33 | | **0.83** | | 0.739 | 0.632 |
| YANCO | 0.8 | | | | | 0.62 | | **0.84** | | 0.837 | 0.829 |

Table 13. A comparison of Pearson correlation (for our best model) vs other works spread across stratified by the sensor network. An asterisk (*) indicates that the time range during which in-situ data was obtained is different from the one we use for training our models. A double asterisk (**) indicates that the work uses region-specific models. A triple asterisk (***) indicates that the work does not ensure a sensor level train/validation split (data points from the same sensor can be present in train and validation).

| | ubRMSE | | | | | | | | | | |
|---|---|---|---|---|---|---|---|---|---|---|---|
| Network | Ours | Bi et al[14]* | Albergel al[15]* | Fang et al[16] | Balenzano et | Collow et | Jing et | Karthikeyan | Abbaszadeh | SMAP | GLDAS |

---

[14] (Bi et al. 2016)
[15] (Albergel et al. 2012) (ECMWF)
[16] (Fang et al. 2021)

|  |  |  |  |  | al[17] | al[18]* | al[19]* | et al[20]** | et al[21]*** |  |  |
|---|---|---|---|---|---|---|---|---|---|---|---|
| AMMA-CATCH | **0.033** |  | 0.036 |  |  |  |  |  |  | 0.072 | 0.04 |
| BIEBRZA_S-1 | **0.076** |  |  |  |  |  |  |  |  | 0.143 | 0.142 |
| FR_Aqui | 0.033 |  |  |  |  |  |  |  |  | 0.122 | **0.032** |
| MAQU | 0.059 | **0.046** |  |  |  |  |  |  |  | 0.061 | 0.063 |
| NAQU | **0.032** | 0.034 |  |  |  |  |  |  |  | 0.041 | 0.046 |
| RSMN | **0.044** |  |  |  |  |  |  |  |  | 0.131 | 0.058 |
| SCAN | 0.052 |  |  | 0.068 |  |  |  |  | **0.047** | 0.086 | 0.06 |
| SMOSMANIA | **0.038** |  | 0.053 |  |  |  |  |  |  | 0.091 | 0.039 |
| SNOTEL | **0.06** |  |  |  |  |  |  |  |  | 0.111 | 0.081 |
| USCRN | 0.054 |  |  | 0.08 |  |  |  |  | **0.04** | 0.082 | 0.06 |
| FORTCOBB | **0.03** |  |  |  |  |  | 0.057 |  | 0.038 | 0.1 | 0.035 |
| KYEAMBA | 0.08 |  |  |  |  |  |  | **0.06** |  | 0.066 | 0.078 |
| LITTLEWASHITA | 0.045 |  |  |  | 0.095 | 0.05 |  |  | 0.041 | **0.035** | 0.103 | 0.06 |
| REMEDHUS | **0.034** |  |  |  | 0.093 |  |  |  |  | 0.073 | 0.041 |
| TWENTE | **0.075** |  |  |  |  |  |  |  |  | 0.123 | 0.105 |
| TXSON | **0.037** |  |  |  | 0.085 |  |  | 0.039 |  | 0.088 | 0.05 |
| YANCO | 0.104 |  |  |  | 0.08 |  | **0.05** |  |  | 0.091 | 0.109 |

Table 14. A comparison of ubRMSE (for our best model) vs other works stratified by the sensor network. An asterisk (*) indicates that the time range during which in-situ data was obtained is different from the one we use for training our models. A double asterisk (**) indicates that the work uses region-specific models. A triple asterisk (***) indicates that the work does not ensure a sensor level train/validation split (data points from the same sensor can be present in train and validation)

Benchmark comparisons for correlation and ubRMSE in Table 13 and Table 14 respectively show that our model performs well across various sensor networks. Even in cases where existing benchmark models perform better, our model trails closely behind. Our model however, performs significantly weaker than most benchmarks on MAQU and RSMN networks in terms of correlation and YANCO in terms of ubRMSE.

---

[17] (Balenzano et al. 2021)
[18] (Collow et al. 2012)
[19] (Jing, Song, and Zhao 2018)
[20] (Karthikeyan and Mishra 2021)
[21] (Abbaszadeh, Moradkhani, and Zhan 2019)

Do note that these are not an exact 1:1 set of comparisons since the time ranges, sampling strategies etc differ for some benchmarks and some of these methods perform an aggregation of in-situ sensor readings falling within a single pixel of their model inputs which simplifies the task to some extent.

*d. Model exploration studies*

1) INPUT SIZE SENSITIVITY ANALYSIS

|  | Validation | | |
| --- | --- | --- | --- |
| **Input size** | **ubRMSE** | **RMSE** | **Correlation** |
| 64x64 | 0.072 | 0.097 | 0.519 |
| 128x128 | 0.068 | 0.094 | 0.554 |
| 256x256 | **0.067** | **0.094** | **0.587** |
| 512x512 | 0.066 | 0.093 | 0.602 |

Table 15. Validation results for the input size sensitivity analysis.

Larger input sizes provide better results since the model has more spatial context and a larger number of model parameters to work with leading to greater representation power, however the improvements decline as we continue to scale up. Considering practical constraints, we went ahead with 256x256 for the input size.

2) INPUT FEATURE ABLATION STUDY

|  | Validation | |
| --- | --- | --- |
| **Feature/Source removed** | **ubRMSE** | **Correlation** |
| **None (Baseline)** | **0.054** | **0.727** |
| DEM | 0.058 (-7.4%) | 0.72 (-1%) |
| Sentinel-1 | 0.056 (-3.7%) | 0.704 (-3.2%) |
| Sentinel-2 | 0.058 (-7.4%) | 0.704 (-3.2%) |
| SoilGrids | 0.064 (-18.5%) | 0.658 (-9.5%) |
| SMAP | 0.057 (-5.6%) | 0.706 (-2.9%) |
| GLDAS | 0.055 (-1.9%) | 0.716 (-1.5%) |
| SMAP + GLDAS | 0.058 (-7.4%) | 0.675 (-7.2%) |

Table 16. Validation results for the feature importance ablation study. Shows the importance of each feature set/source. A '+' in percentage change denotes a relative increase in correlation and a relative decrease in ubRMSE from the baseline.

SoilGrids is the most important source (largest drop from the baseline in terms of performance) for the model followed by Sentinel-2 and Sentinel-1 closely, SMAP next, GLDAS after and finally the DEM as shown in Table 16.

Note that the removal of both the coarse soil moisture inputs drops performance significantly and the drop is close to the drop we see when we remove SoilGrids. Among SMAP and GLDAS, SMAP seems to improve model performance significantly more compared to GLDAS. This could be explained by the fact that GLDAS only provides a 10 cm top level soil moisture estimate instead of a 5 cm estimate that is provided by SMAP.

## 6. Discussion

We developed machine learning based models that assimilate information from different remote sensing and geophysical data sources at varying resolution to produce soil moisture estimates at a nominal resolution of 320m. The result is a trained model applicable in a large variety of settings across the world, and outperforms SMAP/GLDAS baselines and most of the other methods that we compare against in most of the sensor networks. We perform various input sensitivity and ablation studies which provide useful insight into each of the remote sensing sources used in an empirical setting.

Our models are able to capture soil moisture well even in cases where there is high input variability in terms of the terrain and topography. They are not restricted to being applied on homogenous regions only and can be applied in a wide range of settings. This is important since most agricultural farmlands, etc. often have a high amount of variability in a given spatial context (Addis, Klik, and Strohmeier 2015).

We release our models to be used by anyone interested for downstream tasks. Additionally, we curate and release a large scale soil moisture dataset that can be used by others to train and evaluate remote sensing based soil moisture models with ease.

Looking forward, we would like to extend our approaches to perform temporal modelling and add in temporal input sources such as precipitation etc which would strengthen our models further and improve performance. We are also interested in applying more advanced self/semi supervised methods to perform the task at hand to improve model generalization to unseen regions and scale better.

# 7. Acknowledgements

We would like to thank Prof Muddu Sekhar, Christopher H Van Arsdale, Kevin James McCloskey and Rob von Behren for helping us review the paper in whole and Nikhilesh Kumar for providing invaluable suggestions throughout the experimental process.

*a. Data acknowledgements*

1) The USDA Agricultural Research Service, Grazinglands Research Laboratory, El Reno, Oklahoma provided FORTCOBB and LITTLEWASHITA data.
2) The OzNet hydrological monitoring network provided YANCO and KYEAMBA data.
3) The University of Twente via DANS provided TWENTE data.
4) The University of Texas, Austin provided TXSON data via the Texas Data Repository.

This Work has been accepted by the Journal of Hydrometeorology - https://journals.ametsoc.org/view/journals/hydr/24/10/JHM-D-22-0118.1.xml (Batchu et al. 2023). The AMS does not guarantee that the copy provided here is an accurate copy of the Version of Record (VoR).

# 8. Data Availability Statement

We release the dataset we created and used in the paper along with our trained models at https://github.com/google-research/google-research/tree/master/soil_moisture_retrieval.

## APPENDIX

### Sentinel-2 normalization scheme

We use the following non-linear normalization scheme where normalization is applied in the log space which ensures that we have a broader dynamic range for non-cloudy imagery post normalization.

**Formula**:

$$\text{scaled\_band} = e^{\left(\frac{(\log(\text{original\_band} \cdot 0.005 + 1.0) - \log\_mean)}{\log\_std}\right)} \cdot 5.0 - 1.0$$

$$\text{normalized\_band} = \left( \frac{\text{scaled\_band}}{\text{scaled\_band} + 1.0} \right) * 255.0$$

Where the log_mean and log_std for each of the bands are as specified below. These were obtained by computing statistics over the United States.

| Source | Band | log_mean | log_std |
|---|---|---|---|
| Sentinel-2 | B2 | 1.7417268007636313 | 2.023298706048351 |
| | B3 | 1.7261204997060209 | 2.038905204308012 |
| | B4 | 1.6798346251414997 | 2.179592821212937 |
| | B5 | 1.7734969472909623 | 2.2890068333026603 |
| | B6 | 2.289154079164943 | 2.6171674549378166 |
| | B7 | 2.382939712192371 | 2.773418590375327 |
| | B8 | 2.3828939530384052 | 2.7578332604178284 |
| | B11 | 2.1952484264967844 | 2.789092484314204 |
| | B12 | 1.554812948247501 | 2.4140534947492487 |

Table 17. Normalization statistics for Sentinel-2 bands.

### Sentinel-1 anchored dataset validation/test statistics

We provided statistics for the Sentinel-1 anchored dataset in Section 2c2. But to ensure that our validation/test splits are truly IID, we compute the same set of statistics on each of the splits.

1) VALIDATION SPLIT

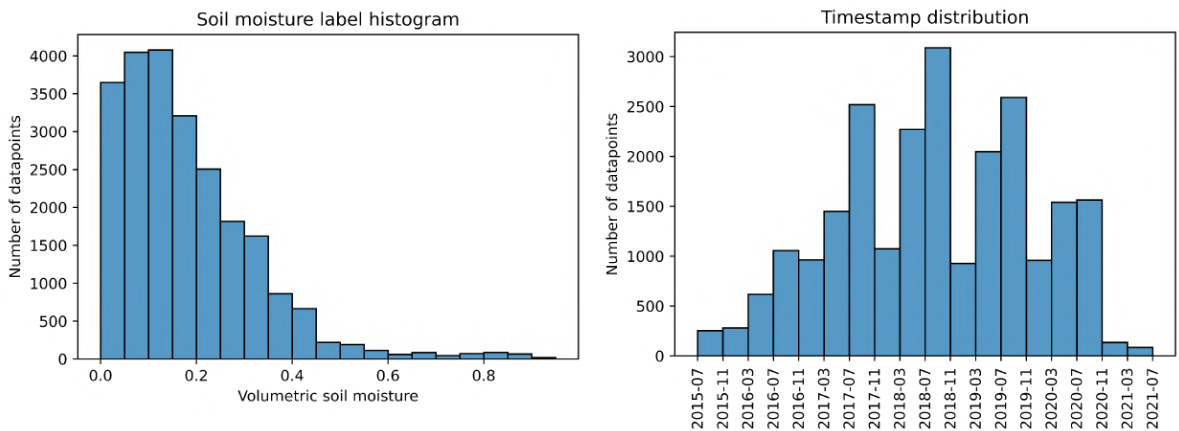

Fig. 16. *Left*: Volumetric soil moisture label distribution. *Right*: Timestamp distribution on the validation split of the dataset.

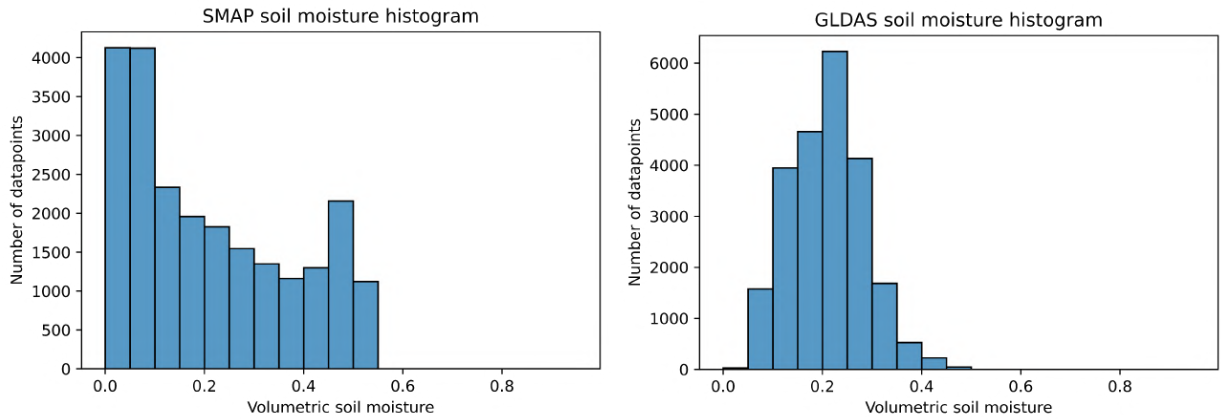

Fig. 17. Distribution of coarse soil moisture products. *Left*: SMAP volumetric soil moisture. *Right*: GLDAS volumetric soil moisture on the validation split of the dataset.

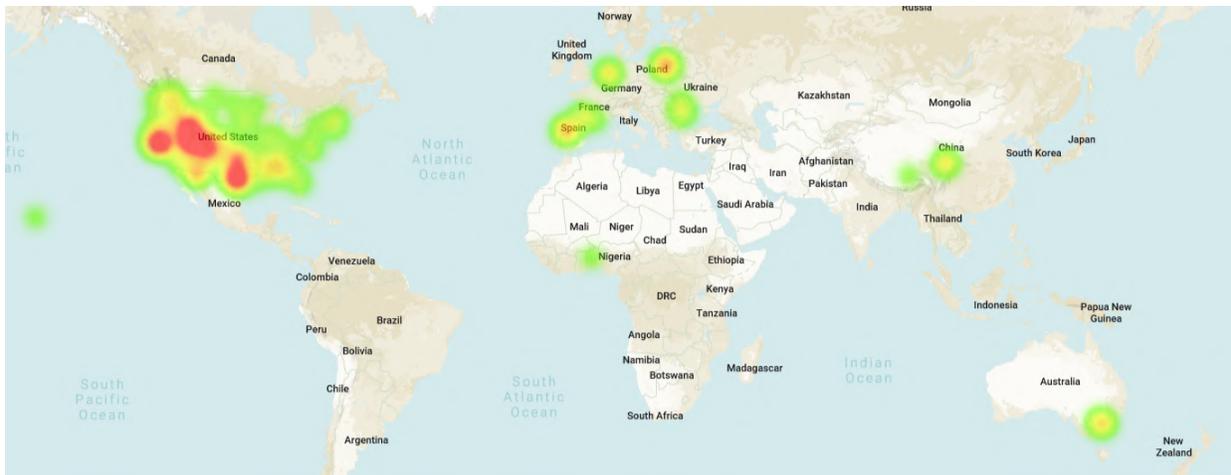

Fig. 18. A heatmap of in-situ sensor locations present in the validation split of the dataset.

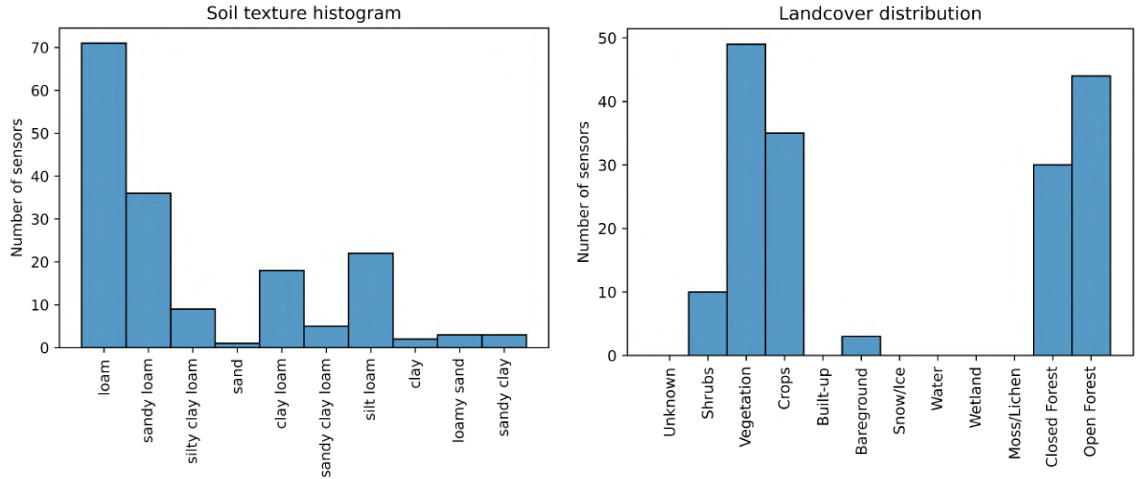

Fig. 19. *Left*: USDA based soil texture distribution. *Right*: Land cover distribution derived from the Copernicus Land Cover Map on the validation split of the dataset.

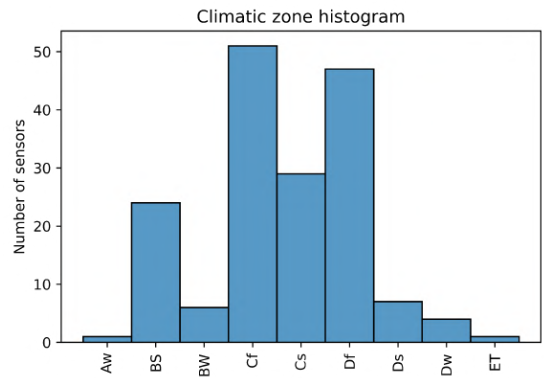

Fig. 20. Köppen climate zone distribution on the validation split of the dataset. We use the first two identifiers of the Köppen classification only in order to group similar climate zones together.

2) TEST SPLIT

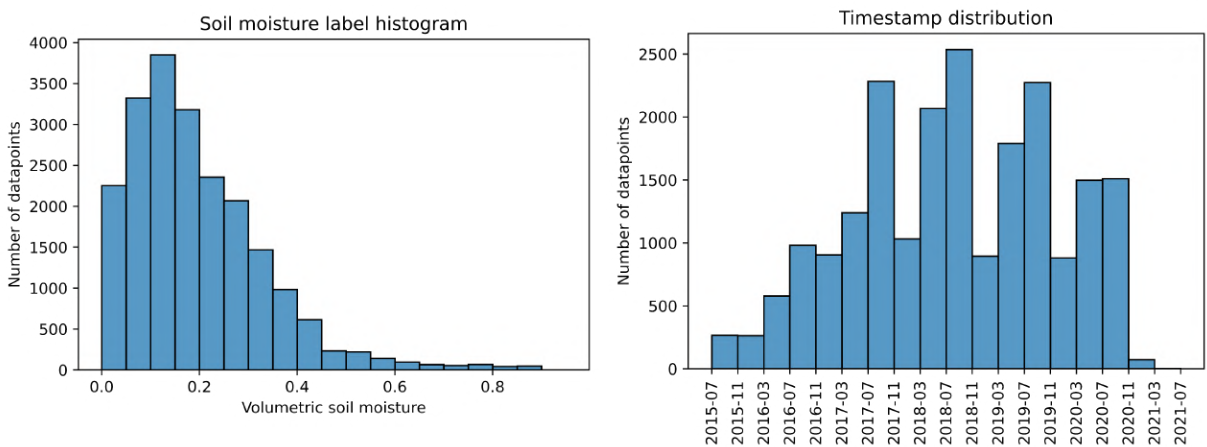

Fig. 21. *Left*: Volumetric soil moisture label distribution. *Right*: Timestamp distribution on the test split of the dataset.

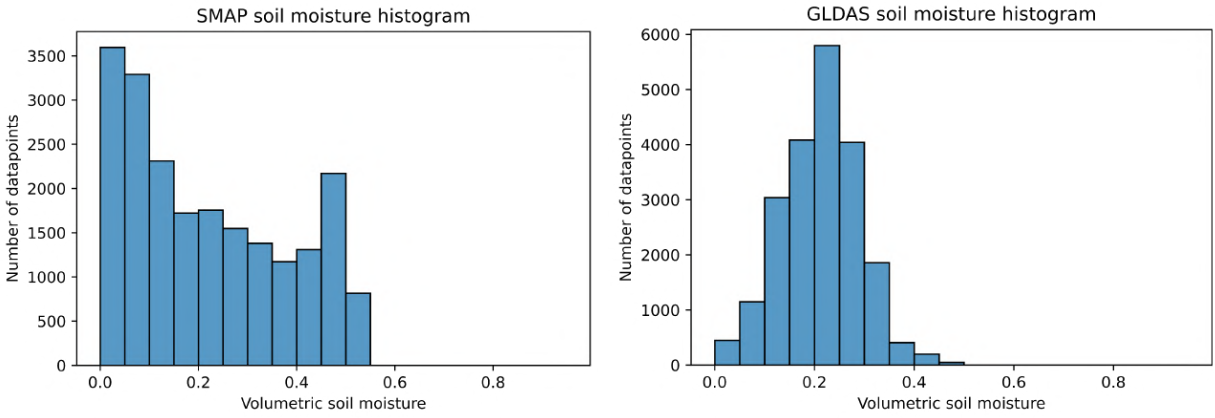

Fig. 22. Distribution of coarse soil moisture products. *Left*: SMAP volumetric soil moisture. *Right*: GLDAS volumetric soil moisture on the test split of the dataset.

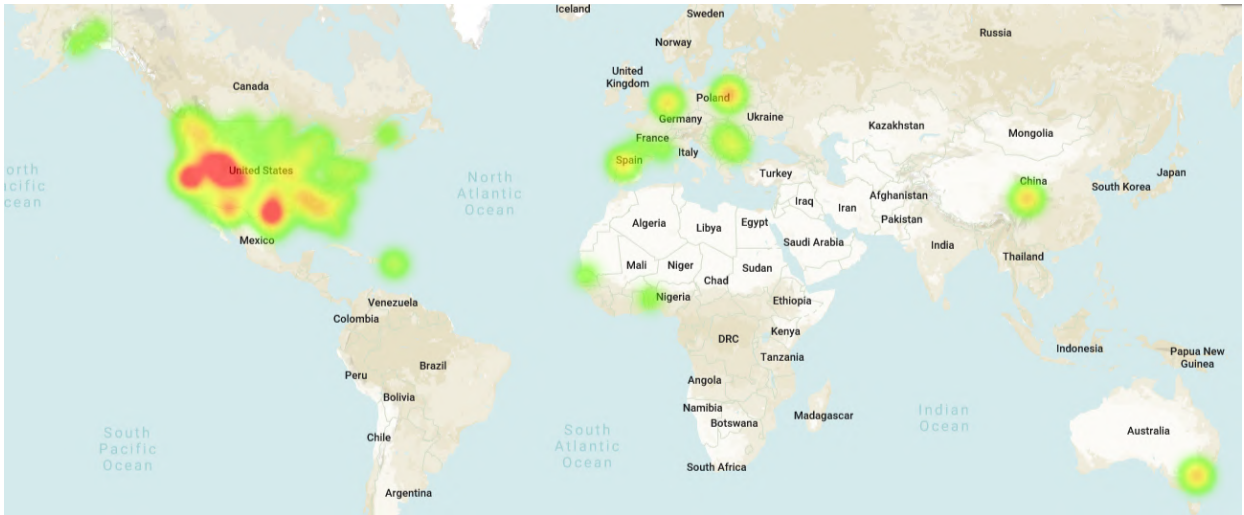

Fig. 23. A heatmap of in-situ sensor locations present in the test split of the dataset.

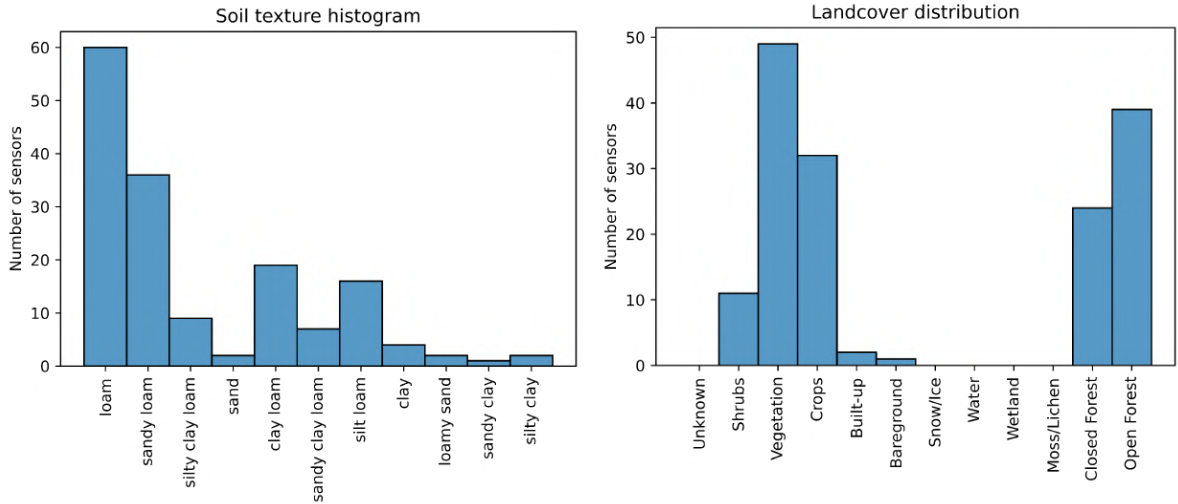

Fig. 24. *Left*: USDA based soil texture distribution. *Right*: Land cover distribution derived from the Copernicus Land Cover Map on the test split of the dataset.

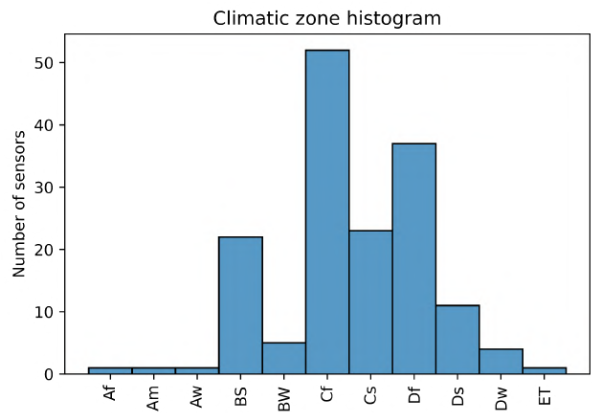

Fig. 25. Köppen climate zone distribution on the test split of the dataset. We use the first two identifiers of the Köppen classification only in order to group similar climate zones together.

From the plots above in Figures 16, 17, 18, 19, 20 and Figures 21, 22, 23, 24, 25, we observe that the distribution of data in our validation and test is similar to our train distribution. This validates that our splits are indeed IID.

**Samples from the Sentinel-1 anchored dataset**

| Sentinel-2 RGB | Sentinel-1 VV | Sentinel-1 VH | SoilGrids | Soil moisture label |
|---|---|---|---|---|
| 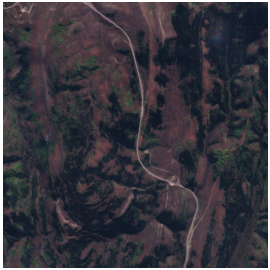 | 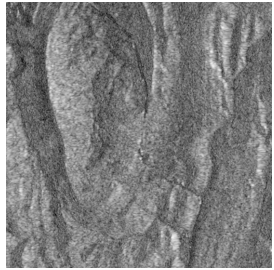 | 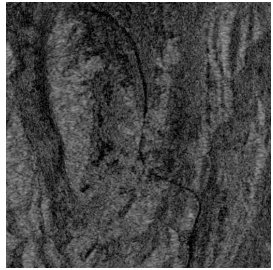 | 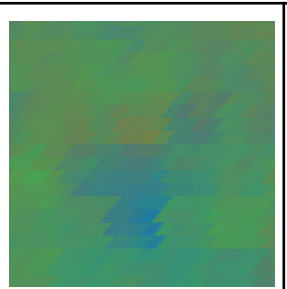 | 0.264 |
| 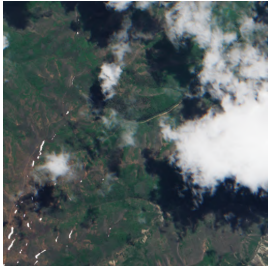 | 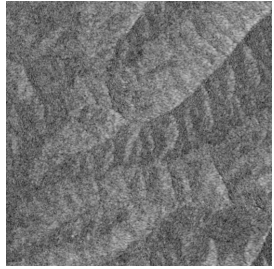 | 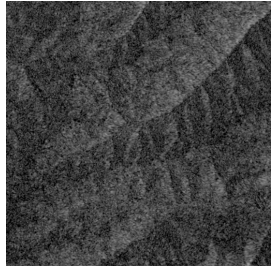 | 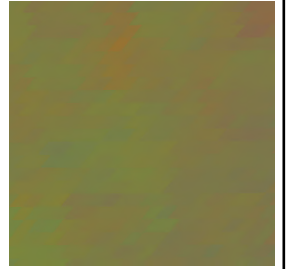 | 0.244 |
| 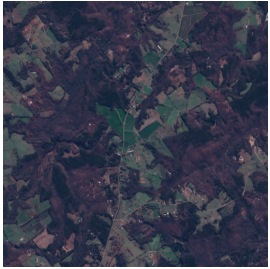 | 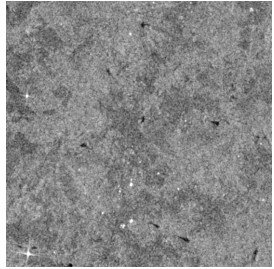 | 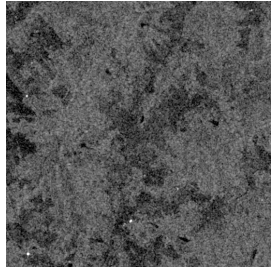 | 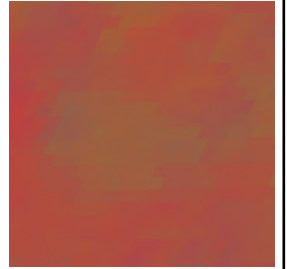 | 0.203 |
| 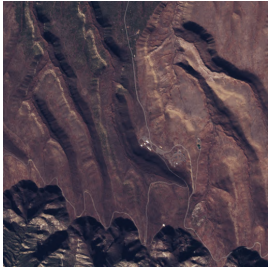 | 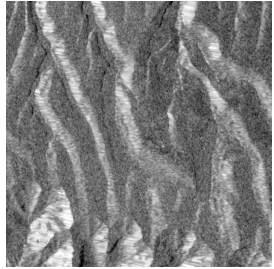 | 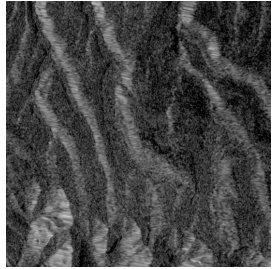 | 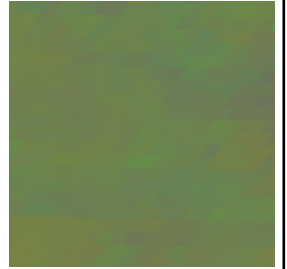 | 0.111 |
| 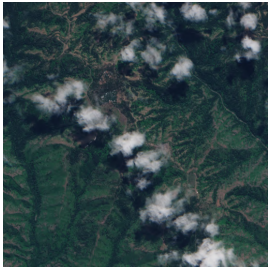 | 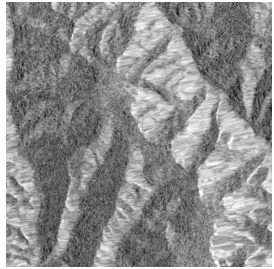 | 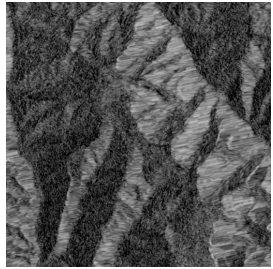 | 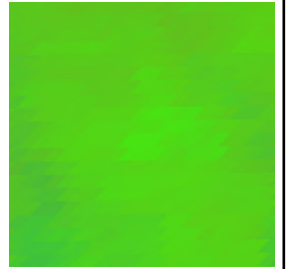 | 0.141 |

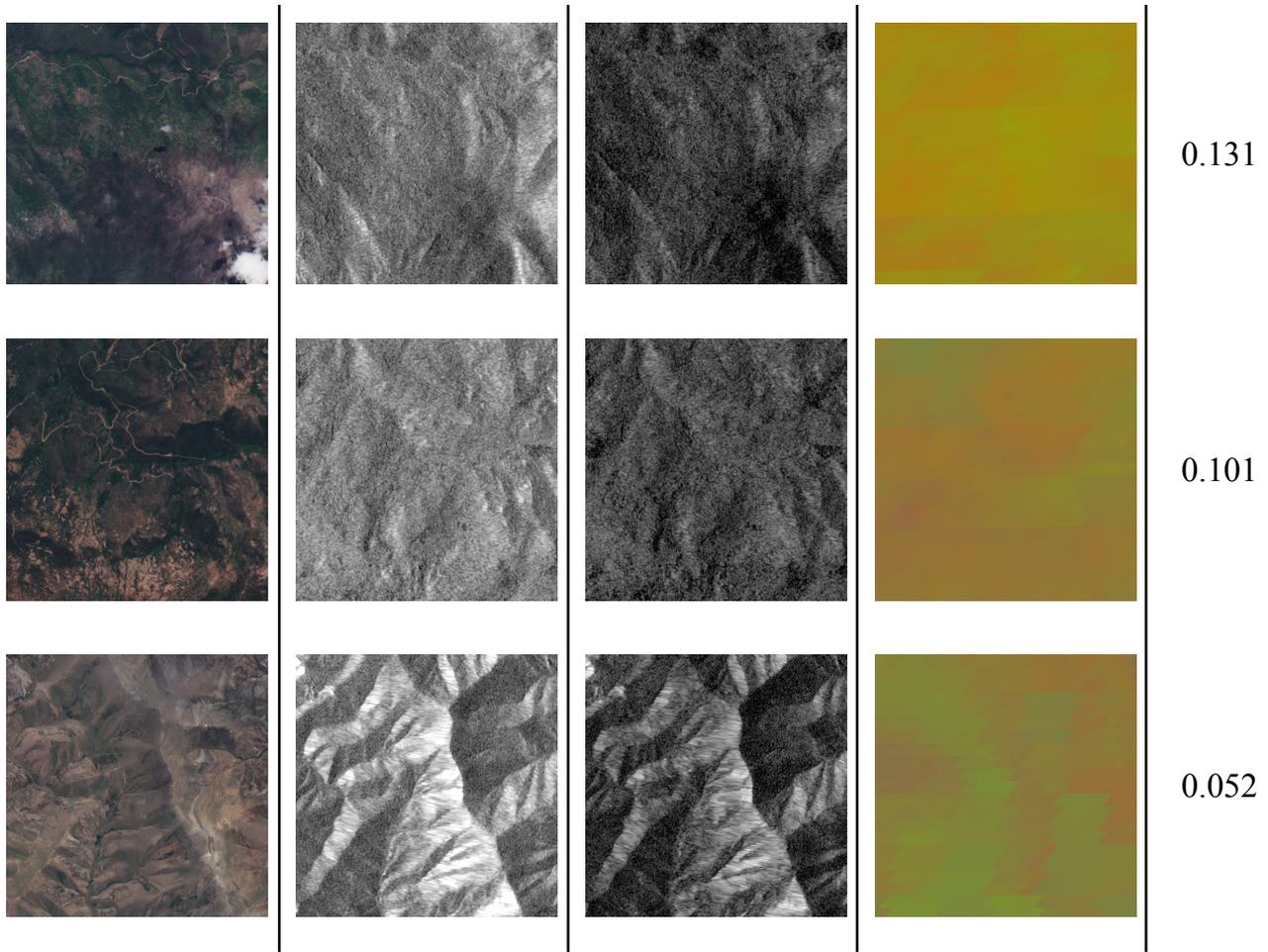

Fig. 26. Randomly selected inputs and labels from the dataset. The SoilGrids image is a false RGB image generated from the [sand, silt, clay] bands for R,G,B respectively.

**Experiments with additional model inputs**

The vegetation indices - Normalized Difference Vegetation Index (NDVI) and Enhanced Vegetation Index (EVI) are commonly used in soil moisture estimation methods. Similarly, the MODIS Leaf Area Index (LAI) is also a widely used input.

We tried to incorporate these into our models - computed vegetation indices from Sentinel-2 added to the high res inputs and MODIS LAI added to the low res inputs. However, we noticed that they didn't improve the performance of our models and even dropped performance slightly as shown in Table 18.

|   | Validation | |
| --- | --- | --- |
| **Feature/Source added** | **ubRMSE** | **Correlation** |
| None (Baseline) | **0.054** | **0.727** |

| EVI/NDVI | 0.0566 | 0.714 |
| MODIS LAI | 0.0567 | 0.714 |

Table 18. Validation results for the feature importance study on additional model inputs. A '+' in percentage change denotes a relative increase in correlation and a relative decrease in ubRMSE.

**Experiments and results on an Sentinel-1 + SMAP (Active-Passive) model**

We also explore an active-passive microwave only model, we use Sentinel-1 + DEM (active microwave source) and SMAP (passive microwave source) as inputs to a model and compare the performance to the rest of our models/baselines. We notice that this mostly performs better than either of SMAP or Sentinel-1 + DEM individually, however significantly falls short in comparison to our best model which uses optical imagery and soil properties and coarse GLDAS soil moisture estimates along with these inputs.

| | Validation | | |
|---|---|---|---|
| **Experiment** | **ubRMSE** | **RMSE** | **Correlation** |
| SMAP | 0.097 | 0.144 | 0.638 |
| Sentinel-1 + DEM | 0.073 | 0.099 | 0.474 |
| Sentinel-1 + DEM + SMAP (Active-passive) | **0.064** | **0.093** | **0.617** |
| **Full model (For comparison)** | | | |
| Sentinel-1 + DEM + Sentinel-2 + SoilGrids + SMAP + GLDAS | 0.054 | 0.085 | 0.727 |

Table 19. Validation results on the Sentinel-1 anchored dataset for the active-passive model compared with SMAP baselines and the best performing model.

**Experiments and results on an SMAP anchored dataset**

We create an SMAP anchored ISMN dataset similar to the Sentinel-1 anchored ISMN dataset used for training models and validation throughout the paper. The purpose of this SMAP anchored ISMN dataset is to have a dataset to evaluate the SMAP baseline against, given that there are large time differences (~3 days) between the Sentinel-1 anchored ISMN dataset and SMAP overpass times. This SMAP-anchored ISMN dataset was created using the temporal bounds specified in Table 20. This dataset is created to evaluate if anchoring the dataset closer to the SMAP overpass improves the SMAP+GLDAS baseline number. The dataset consists of ~400k data points, this is larger than the Sentinel-1 anchored dataset since SMAP has a shorter revisit interval compared to Sentinel-1 which results in more matches with the in-situ data.

| Source | Temporal bounds for SMAP anchored dataset |
|---|---|
| Sentinel-1 | 7 days |
| Sentinel-2 | 14 days |
| SMAP | 1 hr |
| GLDAS | 6 hrs |
| NASA DEM | 50 years (One time) |
| Soil Grids | 50 years (One time) |
| MODIS LAI | 8 days |

Table 20. Temporal bounds for the SMAP anchored dataset.

We perform a few experiments on this SMAP anchored dataset in addition to our experiments on the Sentinel-1 anchored dataset to see how the SMAP + GLDAS NN baseline performance changes depending on the data anchoring we choose.

| | Sentinel-1 anchored data (Validation) | | SMAP anchored data (Validation) | |
|---|---|---|---|---|
| Experiment | ubRMSE | Correlation | ubRMSE | Correlation |
| **Baseline** | | | | |
| SMAP + GLDAS NN | 0.061 | 0.663 | 0.0619 | 0.642 |
| **Ours** | | | | |
| Sentinel-1 + DEM + Sentinel-2 + SoilGrids | 0.058 (+4.9%) | 0.675 (+1.8%) | 0.061 (+1.5%) | 0.632 (-1.6%) |
| Sentinel-1 + DEM + Sentinel-2 + SoilGrids + SMAP + GLDAS | **0.054 (+11.5%)** | **0.727 (+9.7%)** | **0.056 (+9.5%)** | **0.703 (+9.5%)** |

Table 21. Validation results on the Sentinel-1 and SMAP anchored datasets. A '+' in percentage change denotes a relative increase in correlation and a relative decrease in ubRMSE. Note that the percentages specified are changes over the corresponding SMAP + GLDAS NN baseline.

Looking at the results in Table 21, there are two sets of comparisons we'd like to make here,
1) Considering the Sentinel-1 anchored and SMAP anchored datasets individually
    a) Our best models outperform their corresponding baselines by a significant margin in both cases.
2) Comparing the SMAP anchored dataset results vs the Sentinel-1 anchored dataset results
    a) On the SMAP+GLDAS NN baseline, there isn't much of a change, in fact the correlation drops a little. This indicates that using an Sentinel-1 anchored dataset for our baseline results in our experiments, ablations and sensitivity analyses does not undermine the performance of the baseline model.
    b) On the model using Sentinel-1 + DEM + Sentinel-2 + SoilGrids (no coarse soil moisture inputs), we notice a decrease in performance - both ubRMSE and correlation. This indicates that anchoring data being used by the model is important (Sentinel-1 anchoring in this case).
    c) On our best model which uses all the inputs, we still notice a slight decrease in performance. When we anchor on Sentinel-1, SMAP has a 3 day time bound whereas when we anchor on SMAP, Sentinel-1 has a 7 day time bound owing to the smaller revisit interval of SMAP when compared to Sentinel-1. The results indicate that anchoring on Sentinel-1 is more meaningful for our models than anchoring on SMAP (although both Sentinel-1 and SMAP are used as inputs in this model). The hypothesis could be either that Sentinel-1 is a more significant input for our models compared to SMAP, the asymmetry in time bounds between Sentinel-1 and SMAP based on the anchoring or the sampling of labels could be slightly different in both cases.